\documentclass{aa}
\usepackage{psfig}
\usepackage[dvips]{graphics}

\usepackage{graphicx,natbib}
\bibpunct{(}{)}{;}{a}{}{,}

\begin{document}

\title{ISOGAL: A deep survey of the obscured inner Milky Way with ISO at 7$\mu$m and 15$\mu$m and with DENIS in the near-infrared.
\thanks{This is paper no. 20 in a refereed journal
based on data from the ISOGAL project}\fnmsep
\thanks{Based on observations with ISO, an ESA
project with instruments funded by ESA Member States (especially the
PI countries: France, Germany, the Netherlands and the United Kingdom)
and with the participation of ISAS and NASA} 
\thanks {Based on observations collected at the European Southern
Observatory, La Silla Chile.}
}

\author{
A. Omont\inst{1}
\and  G.F. Gilmore\inst{2}
\and  C. Alard\inst{3,1}
\and  B. Aracil\inst{1}
\and  T. August\inst{1}
\and  K. Baliyan\inst{5}
\and  S. Beaulieu\inst{2,32}
\and  S. B\'egon\inst{1}
\and  X. Bertou\inst{1}
\and  J.A.D.L. Blommaert\inst{4,16}
\and  J. Borsenberger\inst{1,24}
\and  M. Burgdorf\inst{4,29}
\and  B. Caillaud \inst{1}
\and  C. Cesarsky\inst{12}
\and  A. Chitre\inst{1,26}
\and  E. Copet\inst{17}
\and  B. de Batz\inst{3}
\and  M.P. Egan\inst{11}
\and  D. Egret\inst{30}
\and  N. Epchtein\inst{13}
\and  M. Felli\inst{8}
\and  P. Fouqu\'e\inst{18}
\and  S. Ganesh\inst{1,5}
\and  R. Genzel\inst{25}
\and  I.S. Glass\inst{14}
\and  R. Gredel\inst{31}
\and  M.A.T. Groenewegen\inst{16}
\and  F. Guglielmo\inst{1}
\and  H.J. Habing\inst{7}
\and  P. Hennebelle\inst{6}
\and  B. Jiang\inst{22}
\and  U.C. Joshi\inst{5}
\and  S. Kimeswenger\inst{19}
\and  M. Messineo\inst{7}
\and  M.A. Miville--Desch\^enes\inst{6}
\and  A. Moneti\inst{1}
\and  M. Morris\inst{20}
\and  D.K. Ojha\inst{9,1}
\and  R. Ortiz\inst{7,23}
\and  S. Ott\inst{4,28}
\and  M. Parthasarathy\inst{27}
\and  M. P\'erault\inst{6}
\and  S.D. Price\inst{11}
\and  A.C. Robin\inst{15}
\and  M. Schultheis\inst{1}
\and  F. Schuller\inst{1}
\and  G. Simon\inst{3}
\and  A. Soive\inst{1}
\and  L. Testi\inst{8}
\and  D. Teyssier\inst{6}
\and  D. Tiph\`ene\inst{17}
\and  M. Unavane\inst{2}
\and  J.T. van Loon\inst{2,10}
\and  R. Wyse\inst{21}
}
\authorrunning{A. Omont, et al}
\titlerunning{ISOGAL-DENIS detection of Intermediate AGB stars in the Galactic
Bulge}
\offprints{A. Omont, omont@iap.fr}

\institute{
Institut d'Astrophysique de Paris, CNRS, 98bis Bd Arago, F-75014 Paris     
\and Institute of Astronomy, Cambridge, U.K.   
\and GEPI, Observatoire de Paris, France  
\and ISO Data Centre, ESA, Villafranca, Spain  
\and Physical Research Laboratory, Ahmedabad, India  
\and Laboratoire de radioastronomie millim\'etrique, Ecole normale sup\'erieure and Observatoire de Paris, Paris, France 
\and Leiden Observatory, Leiden, The Netherlands   
\and Osservatorio Astrofisico di Arcetri, Firenze, Italy  
\and T.I.F.R., Mumbai, India   
\and Astrophysics Group, School of Chemistry \& Physics, Keele University, U.K.  
\and Air Force Research Laboratory. Hanscom AFB, MA, USA   
\and ESO, Garching, Germany    
\and O.C.A., Nice, France    
\and South African Astronomical Observatory, South Africa  
\and Observatoire de Besan\c con, France    
\and Instituut voor Sterrenkunde, K. U. Leuven, Belgium  
\and LESIA, Observatoire de Paris, France  
\and ESO, Santiago, Chile   
\and Institut f\"ur Astrophysik der Leopold-Franzens, Universit\"at Innsbruck, Austria 
\and UCLA, Los Angeles, CA, USA  
\and The Johns Hopkins University, Baltimore MD, USA   
\and Department of Astronomy, Beijing Normal University, Beijing, China 
\and UFES, Vitoria, Brasil
\and UMS-CNRS 2201, Observatoire de Paris, France  
\and MPIE, Garching, Germany
\and Indian Institute of Science, Bangalore, India
\and Indian Institute of Astrophysics, Bangalore, India
\and ESTEC, ESA, Noordwijk, The Netherlands   
\and SIRTF Science Center, California Institute of Technology, Pasadena, CA, USA
\and Observatoire de Strasbourg, France
\and Max-Planck Institut f\"ur Astronomie, Heidelberg, Germany
\and University of Victoria, Canada
}


\voffset 1.0true cm

\date{Received xxxx/ Accepted date}

\sloppy 

\vfill 
\eject

\abstract{The ISOGAL project is an infrared survey of specific regions sampling the Galactic Plane selected to provide information on Galactic structure, stellar populations, stellar mass-loss
and the recent star formation history of the inner disk and Bulge of
the Galaxy. ISOGAL combines 7 and 15~$\mu$m ISOCAM observations -- with a resolution of 6$\arcsec$ at worst -- with DENIS IJK$_{\rm s}$ data to determine the nature of the sources and the
interstellar 
\eject
extinction. We have observed about 16 square degrees with
a sensitivity approaching 10-20mJy, detecting $\sim$10$^5$ sources, mostly
AGB stars, red giants and young stars. The main features of the ISOGAL
survey and the observations are summarized in this paper, together with a brief 
discussion of data processing and quality. The
primary ISOGAL products are described briefly (a full desciption is given in Schuller et al. 2003): viz. the images and the
ISOGAL--DENIS five--wavelength point source catalogue. The main
scientific results already derived or in progress are summarized. These
include astrometrically calibrated 7 and 15~$\mu$m images, determining structures of
resolved sources; identification and properties of interstellar dark clouds; quantification of
the infrared extinction law and source dereddening; analysis of red
giant and (especially) AGB stellar populations in the central Bulge,
determining luminosity, presence of circumstellar dust and mass--loss rate,
and source classification, supplemented in some cases by ISO/CVF
spectroscopy; detection of young stellar objects of diverse types, especially in the inner Bulge with information about the present and recent star formation rate; 
identification of  foreground sources with mid-IR excess. These results are the subject of
about 25 refereed papers published or in preparation.
\keywords{Stars: AGB and post-AGB - Stars: circumstellar matter -
Stars: mass-loss - Stars: formation - Dust: extinction - Infrared: stars - Galaxy: Bulge}}

\titlerunning{ISOGAL: a Galactic survey at 7 and 15~$\mu$m}
\authorrunning{A. Omont et al.}
\maketitle


\section{Introduction}

The ISOGAL project is a multi-wavelength infrared survey at high spatial
resolution of the inner Galaxy, covering the central Galactic Bulge and sampling the obscured
Disk within the Solar circle. The primary scientific aims are to
quantify the spatial distributions of the diverse stellar populations
and their properties in the inner Galaxy, together with the properties
of the warm interstellar medium, and to derive how they are different from those in the vicinity of the solar system.  ISOGAL is based on large-area
broad-band imaging at 7 and 15~$\mu$m with ISOCAM (Cesarsky et al. 1996) on the
ISO satellite (Kessler et al. 1996), and constituted one of the largest programs using
ISO. These mid-infrared data, with a resolution of 6$\arcsec$ at worst, have been 
combined with observations at other wavelengths, mostly in the
near-infrared. To date, most effort has focussed on the association
with complementary DENIS IJK$_{\rm s}$ observations of the central Galaxy
(Epchtein et al. 1997, Simon et al. in preparation).  

Our ISOCAM data (Schuller et al. 2003) are more sensitive by one to four magnitudes, and have
a higher spatial resolution, than the recent survey of the whole Galactic Disk with MSX (see below); they are two orders of magnitude more sensitive, and have
one order of magnitude higher spatial resolution, than IRAS: these improvements allow reliable and nearly complete 
point-source detection down to $\sim$~10~mJy at 15~$\mu$m and
$\sim$~10--20mJy at 7~$\mu$m, depending on crowding and infrared
background (the sensitivity is about a magnitude less in the few fields observed in star-forming regions with very strong background).  The five
wavelengths of ISOCAM + DENIS normally allow reliable determination of
the nature of the sources and of their interstellar reddening. Generally, about 85\% of the ISOCAM sources
are matched with near-infrared sources from the DENIS survey. ISOGAL
data are thus a powerful tool for the analysis of stellar populations in
the most heavily obscured regions of the Milky Way.

While the majority of sources detected only with the deeper 7~$\mu$m
observation are red giants on the upper first ascent branch (RGB), the
most numerous class of ISOGAL 15~$\mu$m sources are Asymptotic Giant
Branch (AGB) stars in the Galactic Bulge and central Disk, with luminosities
above the RGB tip and with evidence for mass-loss in most cases. In addition
to a few foreground main-sequence stars, ISOGAL has also detected a
number of young stars. Most of them have circumstellar dust. These young 
sources are typically stars of a few solar masses in relatively nearby
(1-2 kpc) regions, with a few higher-mass more distant sources. It is
also expected that the ISOGAL sources include a number of as-yet
unidentified peculiar stellar objects of diverse kinds with infrared
excess.
Near-infrared observations, and in particular the DENIS and 2MASS
(Skrutskie et al. 1997, Cutri 1998) surveys, are generally
better able than ISOGAL to detect most normal field stars, especially
red giants and AGB stars with little mass-loss, when the interstellar reddening is
not too high. However, from near-infrared data alone it is extremely
difficult to disentangle interstellar reddening and circumstellar dust
emission. Data at longer wavelengths, which are more sensitive to the
infrared emission from cool circumstellar dust, and are less
affected by interstellar reddening, are required. New mid-infrared space
surveys, such as the present one with ISOCAM, and MSX (Price et
al. 2001, Egan et al. 1996, 1998, 1999), are thus uniquely suited for  carrying out
a census of mass-losing AGB stars and for detecting young stellar objects
in the inner Galaxy. ISOGAL is generally more sensitive than MSX 
by one or two magnitudes at 7--8~$\mu$m, and by three or four  magnitudes at 15~$\mu$m; it has pixels at least three times smaller than MSX. However, MSX has observed the whole Galactic Disk, an area
two orders of magnitude larger than ISOGAL.

In addition to the many stellar sources, the ISOGAL images provide a
high spatial resolution study of the diffuse mid-infrared emission,
yielding a wealth of information about its carriers (PAHs and
dust). ISOGAL allows in particular the identification of dense
globules and filaments, which are opaque even in the mid-infrared
(P\'erault et al. 1996, Hennebelle et al. 2001). The largest of such IR-detected dark clouds were observed by IRAS. ISO and MSX (Egan et al. 1998) showed that they were distinct and sharp edged. The much higher resolution of these experiments permitted them to detect many more dark clouds. The absorption in these clouds also provides a means of measuring the infrared extinction curve along diverse lines of sight.

Many specific ISOGAL results have already been published, including: i)
first imaging results (P\'erault et al. 1996); ii) a study of late-type
giants and long--period and semi--regular variables in Baade's Windows (Glass et
al. 1999, Glass \& Alves 2000, Alard et al. 2001); iii) AGB stars and their mass--loss
rates in the intermediate inner Bulge (Omont et al. 1999); iv) source
counts in specific fields by combining with the ISOCAM observations taken by the SWS dedicated time experiment GPSURVEY (Burgdorf et al. 2000); v) discoveries of
young stellar objects (Testi et al. 1997, Felli et al. 2000, 2002), Schuller (2002), and VLA follow--up observations (Testi et al. 1999); 
vi) properties of dark interstellar globules (Hennebelle et
al. 2001); vii) and subsequent millimetre (Teyssier et al. 2002) and
submillimetre (Pierce-Price et al. 2000) observations. A number of other
detailed analysis papers have been published, submitted for publication or are in
preparation: associations with OH masers (Ortiz et al. 2002), new SiO
masers (Messineo et al. 2002), analysis of specific fields (Ojha et
al. 2003, van Loon et al. 2003, Jiang et al. 2003, Schuller 2002 and Schuller et al. in
prep., Ganesh et al. in prep.), a catalogue of 
luminous infrared sources with cross-identifications  in the central Bulge and Disk (Chitre et
al. in prep.), visible spectroscopy (Schultheis et al. 2002),
near-infrared spectroscopy (Schultheis et al. 2003), ISOCAM CVF
spectroscopy (Blommaert et al. in preparation).

The present paper, accompanying the publication of the Point Source
Catalogue (PSC) of 7 and 15 $\mu$m ISOGAL sources cross-identified with DENIS, aims at
a general presentation of the ISOGAL survey, of its products and their
content, and of its present and expected achievements for the
different classes of topics, fields and objects. A companion paper
(Schuller et al. 2003) contains the Explanatory Supplement to the PSC
 and a detailed analysis of ISOGAL data processing and
quality.

The paper begins with a brief summary of the main features of the ISOGAL
observations, data processing and quality, and products
(Sect. 1,2,3). We then discuss in turn the impact of ISOGAL on the main
topics of interest: interstellar medium, stellar populations, AGB
circumstellar dust and mass--loss, and young stars.

\section{Observations}


\begin{figure*}[htbp]
\begin{center}
\resizebox{18cm}{!}{ \rotatebox{0}{\includegraphics{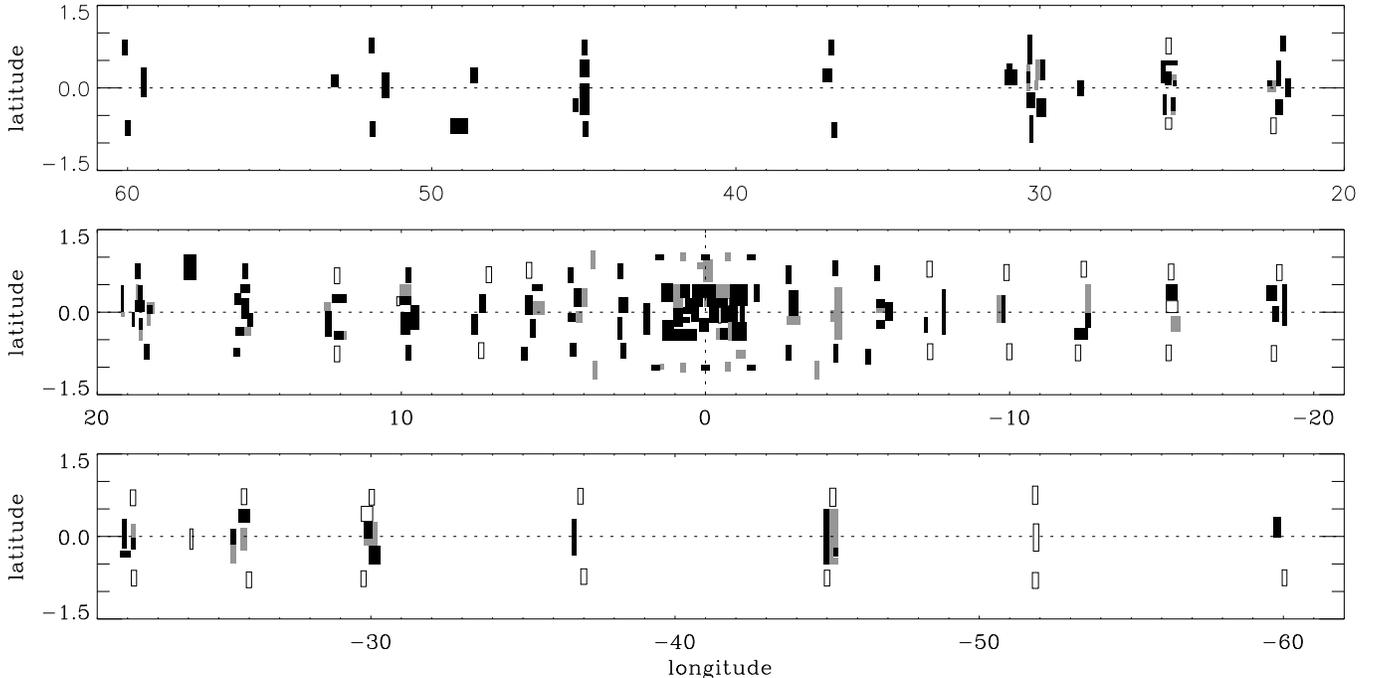}}}
\caption{Galactic map of ISOGAL fields with $\vert \ell
\vert$~$<$~60$^\circ$. Black, grey and open boxes show
fields which have been observed at both 7 \& 15~$\mu$m, at 7~$\mu$m
only and at 15~$\mu$m only, respectively. Twenty--one additional
northern fields are not displayed, at $\ell$~$\approx$~+68$^\circ$,
+75$^\circ$, +90$^\circ$, +98$^\circ$, +105$^\circ$,
+110$^\circ$, +134$^\circ$, +136$^\circ$, \& +138$^\circ$ (see Schuller et al. 2003).}
\end{center}
\label{figure1}
\end{figure*}

The ISOGAL observational program with ISOCAM extended from January 1996 to April
1998, i.e., over practically the whole ISO mission. It is the result of three successive proposals developed as the mission progressed, which 
unfortunately prevented simultaneous 7 and 15~$\mu$m observations in
many fields. The observed regions ($\sim$16 deg$^2$) were distributed
along the inner Galactic Disk, mostly within $\vert \ell
\vert$~$<$~30$^\circ$, $\vert {\it  b} \vert$~$<$~1$^\circ$, to provide a
regular sampling pattern, as shown in Figure 1. The central region
($\vert \ell \vert$~$<$~1.5$^\circ$, $\vert {\it b} \vert$~$<$~0.5$^\circ$)
was almost entirely observed in one wavelength at least, except very
close to the Galactic Center and a few other problematic areas (Fig. 3). These omissions were necessary to avoid very strong sources, which would have
saturated the ISOCAM detectors, causing long--lasting adverse side--effects. 

Thus most of the fields were chosen to avoid strong IRAS sources. Each observation consisted of 
small rasters ($\sim$0.1 deg$^2$), oriented in Galactic
coordinates. 
Standard ISOGAL observational conditions included (see Schuller et al. 2003): broad ISOCAM
filters, LW2 (5-8.5~$\mu$m) and LW3 (12-18~$\mu$m); 6$\arcsec$ pixels;
half-overlapping 3'~x~3' ISOCAM frames (32~x~32 pixels) and short total integration time
(76~x~0.28~s = 21~s) per frame, with overhead time $\approx$~10~s per frame. 

	In order to allow the observation
of active star-forming regions, especially in the central area, the initial
conservative flux-avoidance limit (6~Jy at 12~$\mu$m) was subsequently relaxed by adopting: i) narrower filters (LW6
7.0-8.5~$\mu$m, LW5 6.5-7.3~$\mu$m, LW9 14-16$\mu$m), ii) 3$\arcsec$ pixels, and allowing explicitly for the
fact that the flux of stronger IRAS sources, especially young stellar
objects, is distributed over more than one ISOGAL pixel. In the last
months of the ISO mission, the ISOGAL observations were extended in
several directions, namely in regular sampling of the northern Galactic Disk up
to $\ell$~=~105$^\circ$, in sampling of a few Bulge fields farther from
the Galactic Plane 1$^\circ$\,$<$\,$\vert${\it b} $\vert$\,$<$\,6$^\circ$, in specific
observation of a few star-forming regions
(16$^\circ$~$<$~$\ell$~$<$~138$^\circ$), in repeated and other
verification observations, and in taking CVF spectra (5-16$\mu$m) of eighteen ~3'x3'
fields.

Altogether, 250 hours of observing time were used, which makes ISOGAL one
of the largest ISO programs. Detailed information on the
observation parameters and in particular on the limits of the observed
rasters is available in Table 3 of the ISOGAL Explanatory Supplement
(Schuller et al. 2003) and on the ISOGAL web server ~~~~~~~~~~
(www-isogal.iap.fr/).

Systematic cross-identification with the near-infrared K$_{\rm s}$,~J,~I sources
of the DENIS survey is an integral part of the ISOGAL program. Special
DENIS observations of the ISOGAL fields were performed to provide
early availability of the data (Simon et al. in preparation). We are
at present (early 2003) making available at the CDS Strasbourg data
centre a five-wavelength catalogue of ISOGAL point sources with their
DENIS identification, if available. The latter covers almost the whole observed
area, excluding at present only the few northern ISOGAL fields with
$\delta$~$>$~+2$^\circ$.  The ISO archive also contains LWS data
obtained in parallel mode, for regions of the sky 12' apart, which will provide the spectral energy
distribution of the far-infrared diffuse emission in the ISOGAL field
areas (Lim et al. 2000); but we do not consider these LWS data to be a
part of the ISOGAL program.

\section{Data processing and analysis} 

Reducing ISOCAM data requires a considerable number of operations (see ISOCAM Handbook, Blommaert et al. 2001).
The extensive data processing packages now available handle 
well the usual features of all ISOCAM data: glitches, the dead column,
the time-dependent behavior of the detectors. Additional difficulties
in reducing the ISOGAL data are due to the bright background of highly
structured diffuse emission; the high spatial density of bright
sources, which induce long-lasting remanence effects; our very short
integration times; unavoidable undersampling of the images;  no
redundancy in sky-sampling; and image crowding in the
fields. Therefore, a special reduction pipeline was devised by
C. Alard, S. Ganesh and F. Schuller, to complement the standard treatment applied to the
ISOCAM data as delivered (Schuller et al. 2003). This special pipeline
uses several procedures of the Interactive Analysis System for ISOCAM (CIA) software [jointly developed by the
ISOCAM Consortium and the ESA Astrophysics Division (Ott et
al. 1997, Blommaert et al. 2001
)] which are not implemented in the standard treatment of ISO
data. It also includes sophisticated source extraction, after
regularisation of the point-spread-function (psf), and psf
determination directly from the average ISOGAL data. The details of
the processing of ISOGAL data and the assessment of their present
quality are discussed in a companion paper (Schuller et al. 2003).

The completeness of point source extraction has been systematically
addressed through retrieval of added artificial sources. Such
completeness findings have been complemented and checked by the
results of a few repeated observations (in one case with 3$\arcsec$ pixels,
rather than the typical 6$\arcsec$ pixels, and hence with greatly reduced
crowding), and by comparison with DENIS (or 7~$\mu$m) red giant source
counts. The completeness limit depends on the source density, on the
intensity and the structure of the local diffuse background, and on
the filter used. The sensitivities reached at 7 and 15~$\mu$m for
standard ISOGAL conditions are summarized in Table 1.

\begin{table*} 
\begin{center}
\caption[]{Sensitivities\footnotemark[1] at 7 and 15~$\mu$m for
typical ISOGAL conditions} 
\vspace{0cm}
\begin{tabular}{l c c c c c c c c}\\
\hline
Region\footnotemark[2] & Source & Background & Pixel & Filter & \multicolumn{2}{c}{7
$\mu m$} & \multicolumn{2}{c}{15~$\mu m$}  \\ 
 & density &       &  &   & mag & flux (mJy) & mag & flux (mJy)  \\
\hline
A & low       & weak   & 6\arcsec & broad  &10& 9&8.7& 7 \\
B & high      &moderate& 6\arcsec & broad  & 9&22&8  &12 \\
C & very high & strong & 3\arcsec & narrow & 8.4&35&7  &30 \\
D & high   &very strong& 6\arcsec & narrow & 7.7 & 55 & 6.5  &  45 \\
\hline
\label{table1}
\end{tabular}
\end{center}
\vspace*{-0.5cm}
\begin{small}
\noindent 
\footnotemark[1] Sensitivity limits of ISOGAL sources published in PSC
{\it Version 1}, corresponding approximately to detection completeness
of 50\% (Schuller et al. 2003).\\ 
\footnotemark[2] Typical regions:\\
A Lowest density Bulge fields, $\vert${\it b}$\vert$ $\ge$ 2$^\circ$\\
B Standard Disk fields, $\vert${\it b}$\vert$ $<$ 0.5$^\circ$, $\vert$$\ell$$\vert$ $\le$ 30$^\circ$\\
C Central Bulge/Disk fields, $\vert${\it b}$\vert$ $<$ 0.3$^\circ$, $\vert$$\ell$$\vert$ $\le$ 1$^\circ$\\
D Most active star formation regions such as M16, W51. 
\end{small}  
\end{table*}

As discussed in Schuller et al. (2003) a special care was taken in order to discard from the source lists possible remnant ``ghosts'' in the pixels which have seen a strong source in previous observation frames, and which may appear as normal sources in the images.

The statistical reliability of the final extracted point sources was checked through
repeated observations of a few fields, and by confirmation of their reality by 
 their detection in another ISOCAM or DENIS band. Such checks
have shown that there are still a few unconfirmed sources,
possibly spurious, among the weak ones with poor quality flags (see Schuller et al. 2003). These could be
related to incomplete correction of remnant
``ghosts'', background noise, straylight and other artefacts. Therefore, we 
limit the discussion to sources brighter than a lower limit, generally greater than $\sim$10~mJy in both
bands, depending on the confusion and background noise in each field (see Schuller et al. 2003 and Table 1). Nevertheless we suggest that users should carefully question the
reality of faint sources with poor quality flags not detected in at least two ISOCAM or DENIS bands.

Photometric accuracy is estimated from repeated observations and
retrieval of added artificial sources. Typical rms is $\sim$0.2~mag,
though higher for the faintest sources. A systematic bias in the
absolute calibration of $\sim$0.1-0.2~mag is still possible because,
in particular, of the difficulty of correcting transient effects. Both
dispersion and bias increase close to the confusion limit in crowded
fields.

Cross-identifications of LW2 and LW3 sources, either between each other or
with DENIS sources, provide the main information for discussing the 
nature of the sources and their properties. Cross-identification efficiency can also
be quite useful for assessing data quality. We applied routine
standard procedures for 7\,$\mu$m--15\,$\mu$m and DENIS--ISOCAM
cross-identifications (Schuller et al. 2003). The good quality of the
pointing of ISO and of the correction of the ISOCAM field distortions
permits, after optimisation of a small translation of each 
field, the rms residual of the nominal separation  of
matched sources to be reduced to $\sim1-2"$. This is thus the order of magnitude of
the accuracy of the astrometry of the ISOGAL sources not matched with
DENIS sources (except in the few fields without DENIS
observations). The other ISOGAL sources, the large majority, being
matched with DENIS sources, benefit from the excellent DENIS
astrometry ($<$~$\sim$0.5$\arcsec$).

In conclusion, although we hope that significant improvements will
still be possible in the future, the present quality of the data is
sufficiently acceptable, as concerns reliability, completeness, photometric
accuracy and astrometry of the sources, to enable systematic
scientific analysis.

\section{ISOGAL products}

\begin{figure*}[htbp]
 \centerline{
   \resizebox{8cm}{!}{\includegraphics{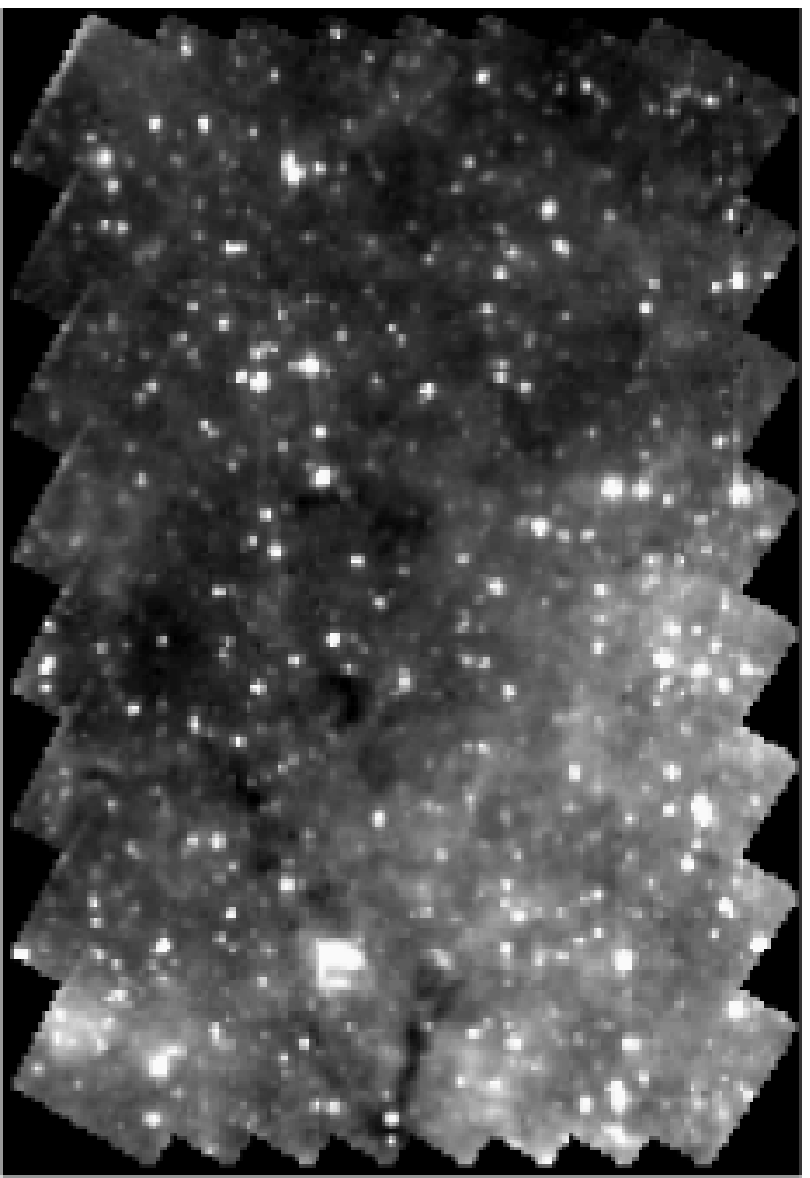}}
   \resizebox{2cm}{!}{\includegraphics{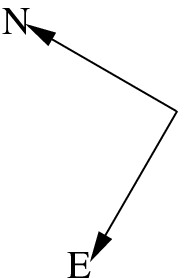}}
   \resizebox{8cm}{!}{\includegraphics{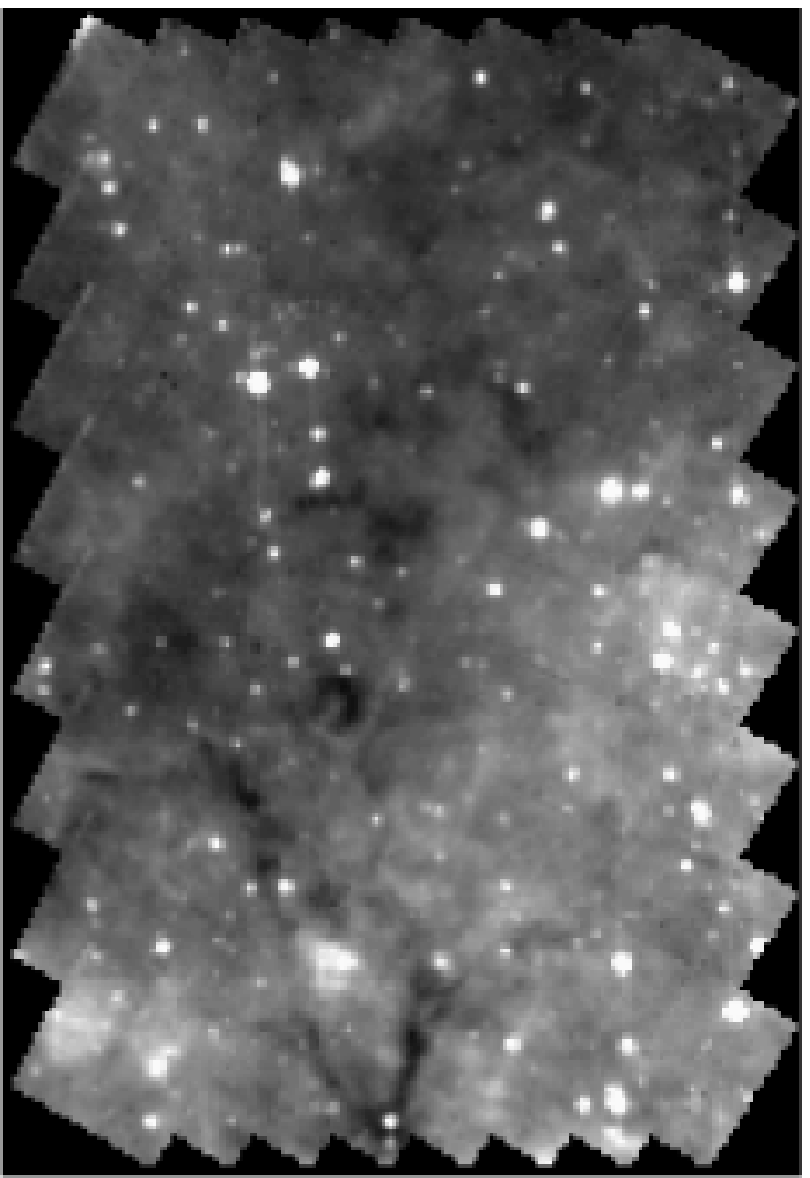}}
 }
\caption{Typical images of ISOGAL rasters: left panel- at 7~$\mu$m (LW2 filter)
and right panel- at 15~$\mu$m (LW3 filter), of field FC+00733+00015, centered
at $\ell$~=~+7.33$^\circ$, {\it b}~=~+0.15$^\circ$ (0.10~deg. x
0.16~deg.) (see Schuller et al. 2003 for the naming convention of ISOGAL fields). Patchy or filamentary infrared dark clouds are visible,
particularly in the centre and at the bottom of the 15~$\mu$m image}
\end{figure*}

\begin{figure*}[htbp]
\begin{center}
\resizebox{16cm}{!}{\includegraphics{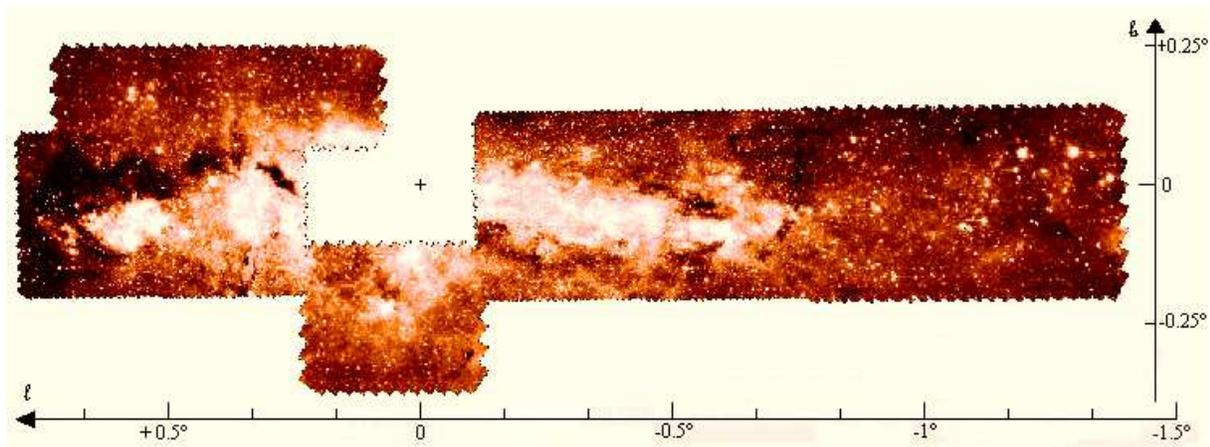}}
            \caption{Mosaic of ISOGAL images at 7~$\mu$m (narrow filter LW5), mostly
3$\arcsec$ pixels, in the neighbourhood of the Galactic Center (from Ganesh
et al in preparation, image also available at
http://antwrp.gsfc.nasa.gov/apod/ap000629.html). Spectacular dark
condensations are visible, especially the string at
$\ell$~$\approx$~0.3$^\circ$--0.8$^\circ$, between the Galactic
Center and Sgr B2 ($\ell$ = 0.67$^\circ$, {\it b=} = $-$0.04$^\circ$).}
\end{center}
\label{figure3}
\end{figure*}


\begin{figure*}[htbp]
 \centerline{
   \resizebox{8cm}{!}{\includegraphics{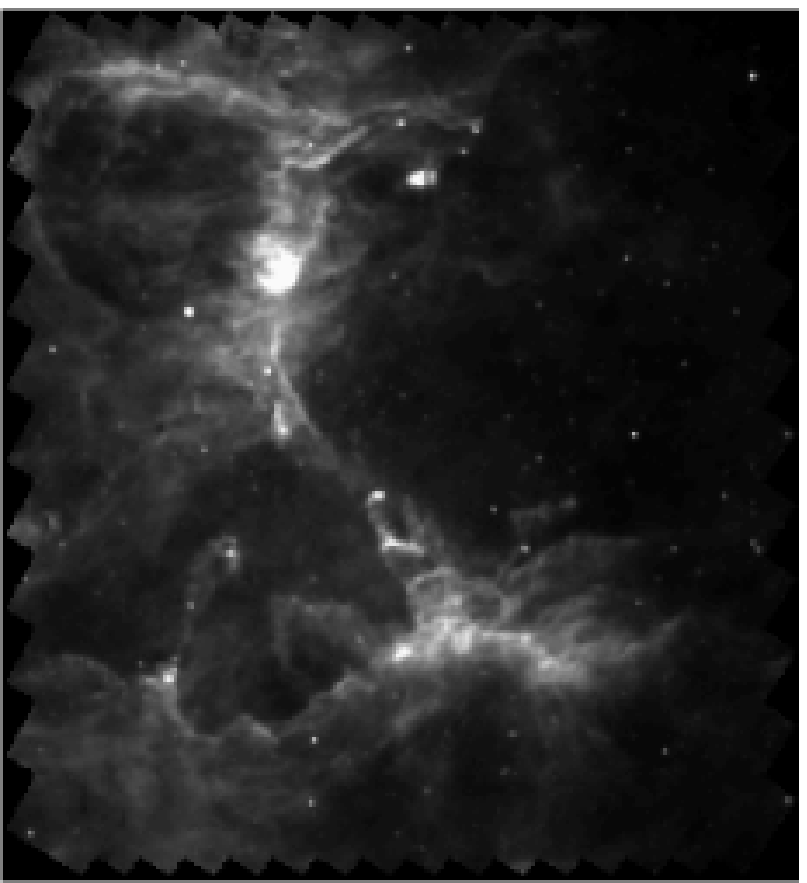}}
   \resizebox{2cm}{!}{\includegraphics{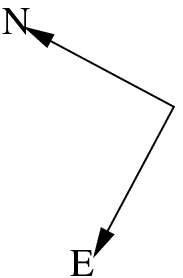}}
   \resizebox{8cm}{!}{\includegraphics{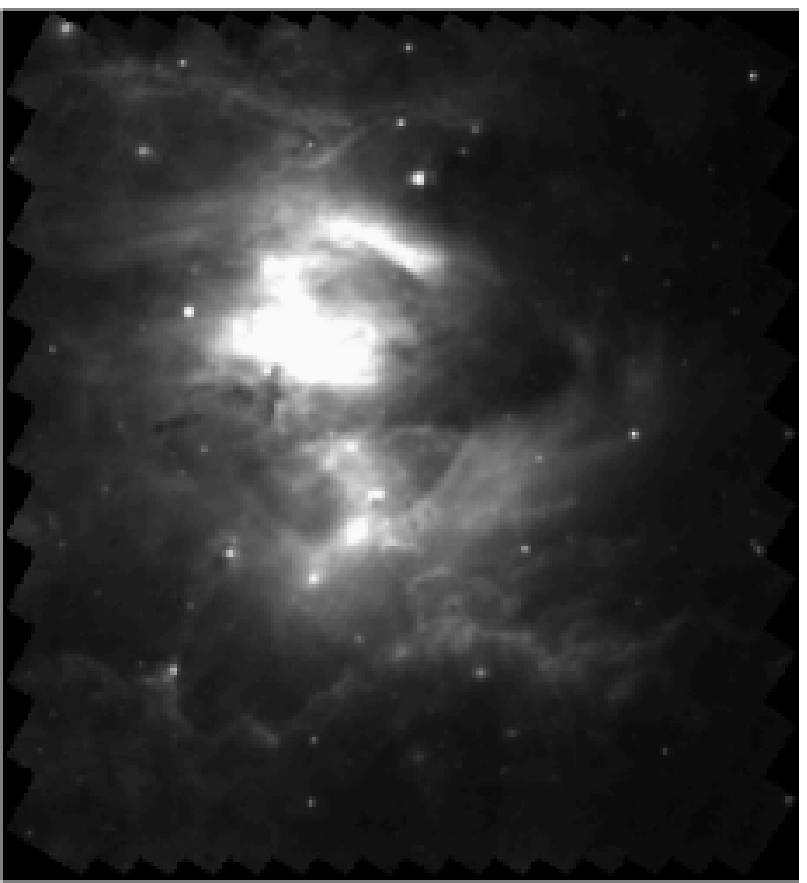}}
}
\caption{Images of the star-forming region M16 (Eagle Nebula). Left panel shows the
7~$\mu$m image, in the narrow LW6 filter, displaying the extraordinary
filamentary structure of the PAH emission. On the other hand, in the right panel, the
15~$\mu$m image, with the narrow LW9 filter which includes the 15.6
$\mu$m Ne III line, displays a completely different structure centered
on the H II region. A composite of both images is available at
http://antwrp.gsfc.nasa.gov/apod/ap010914.html}
\end{figure*}


In addition to the standard data available in the ISO archives, the
ISOGAL Collaboration is making available more elaborate products,
especially for stellar sources, from the data processing described
above. A first version is about to become available through CDS/VizieR. It is intended
to improve it in the future and to complete it with CVF 4--16~$\mu$m
spectra of eighteen ~3'x3' ISOCAM fields. The present data, the images and
the multi--wavelength point source catalogue (PSC), are described in
detail in their Explanatory Supplement in Schuller et al. (2003).

The ISOGAL images available at present are products of the ISO/CIA
processing ((Ott et al. 1997, Blommaert et al. 2001
). They are mosaiced images with the so-called ``inversion'' treatment
of the time dependence of the detector signal. They still contain some  obvious defects and it is intended to
publish improved images soon with a better correction of transient and remnant effects (Miville--Desch\^enes et al. 2000 and in
preparation). However, the astrometry of the present images has been
improved by calibration through DENIS astrometry in most fields; it is thus
significantly better than the standard ISOCAM astrometry. An image is
provided for every reasonably good quality ISOGAL observation
($\ell$~x~{\it b} raster with given filter and pixel-field-of-view,
corresponding to a given ISO Observation Number, ``ION''), amounting to a total of 384
images. Fits images are available at the web page: ~~~~~~~~~~~~~~~~~~
http://www-isogal.iap.fr/Fields/index\_tdt.html. In addition, specific lists are provided to identify and
discard the spurious sources present in the images. Two-colour
images are also provided, by combining 7 and 15~$\mu$m data for the
fields where both are available 

(http://www-isogal.iap.fr/Fields/).

The ISOGAL PSC at present gives magnitudes, I, J, K$_{\rm s}$, [7], [15],  at five wavelengths (0.8, 1.25, 2.15, 7 \& 15~$\mu$m) with DENIS providing  I,J,K$_{\rm s}$
associations when available. The PSC contains ~105000 sources, of
which about half have 7-15~$\mu$m associations and 78\% have DENIS
associations. Since, in many cases, for observational reasons, the
rasters observed at 7 and 15~$\mu$m do not exactly coincide, the PSC
is organised in classes of ``fields'' derived from the observed
rasters : fields observed both at 7 and 15~$\mu$m (163 fields); fields
observed only at either 7 or 15~$\mu$m (43 and 57 fields,
respectively). Each ``field'' covers a rectangle in  $\ell$~x~{\it b} fully observed 
and sources too close to the ``saw-tooth'' edges are discarded. When
several observations at the same wavelength (7 or 15~$\mu$m) exist for a sky
area, only one is used  in the present catalogue. Quality flags
are provided for each source at each wavelength, as well as for source
associations, and only sources with a reasonable quality are
published (see Schuller et al. 2003). 
A number of spurious sources, visible in the present images, are not
included in the PSC because they are suspected to be artefacts, in
particular possible remnants in pixels which have previously seen
bright sources. For each field, such spurious sources are published in
special tables (Schuller et al. 2003).
Additional unpublished information on ISOGAL sources can be
requested for special purposes from the ISOGAL PI (A.Omont).

\section{Interstellar Medium}

\subsection{Images}

        \begin{figure}[]
        \centerline{
   \resizebox{9cm}{!}{\includegraphics{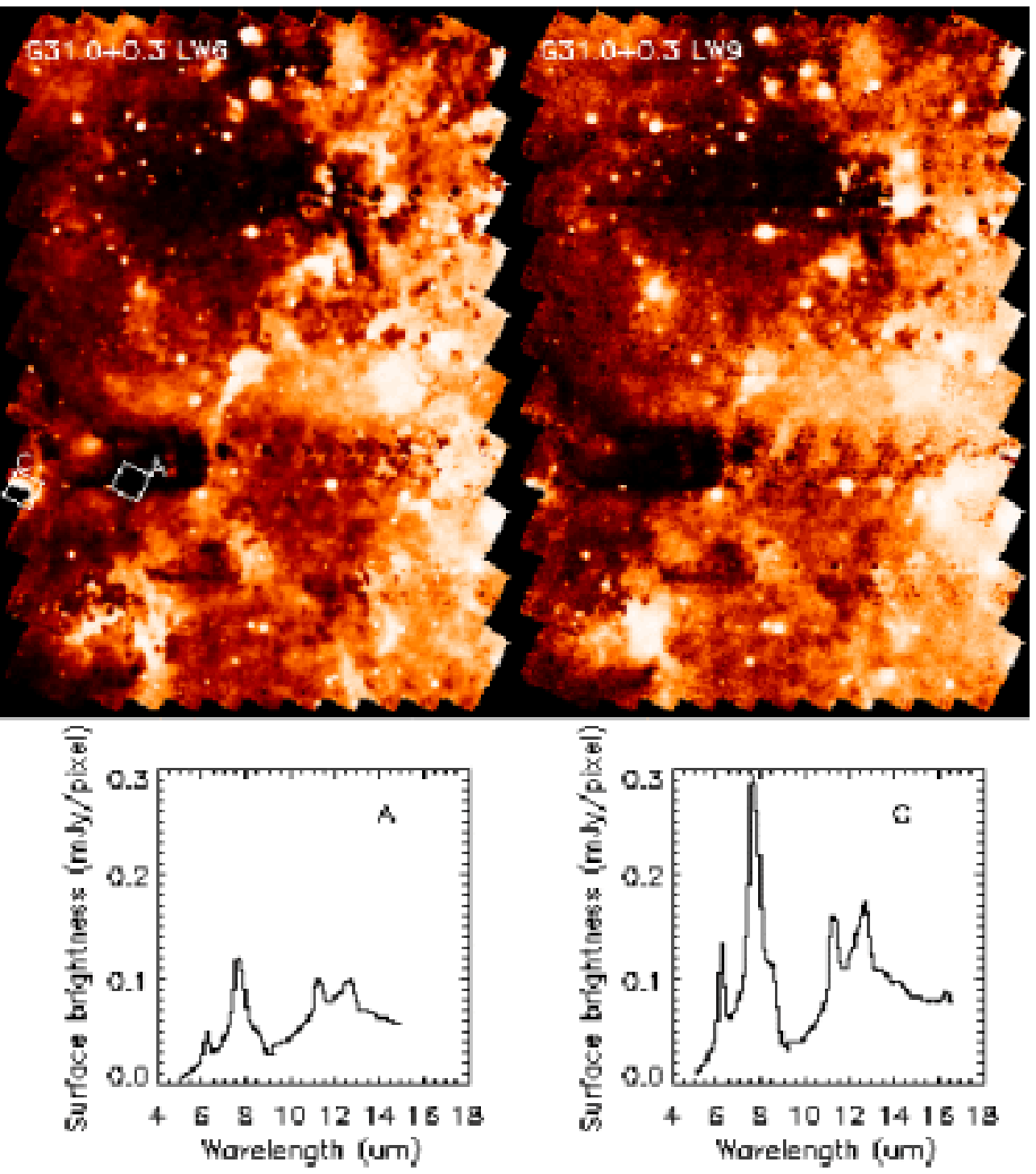}}}
	\caption{Typical CVF spectra of the diffuse
interstellar emission  at two locations A and C, marked on the image (P\'erault et al. in
preparation): A) line of sight of an infrared ``dark'' cloud absorbing the background diffuse emission; C) typical region with moderately bright emission. The strong PAH features between 6
and 13 $\mu$m predominate, with some continuum emission mainly at long wavelength, probably from small grains}
        \label{figure5}
        \end{figure}

The high spatial resolution information on
diffuse mid-infrared emission in ISOGAL images provides a wealth of
information about its carriers (polycyclic aromatic hydrocarbons -- PAHs -- and dust), as well as about the
interstellar radiation which excites them. All ISOGAL images display a
rich structure of such diffuse emission, especially in or close to regions of
star formation (Figs. 2--5). From ISOCAM studies of the interstellar medium in our
Galaxy and others, it is well-established that the most
important contributors to the emission in the diverse 7~$\mu$m filters
are the PAH bands. This is exemplified in the CVF spectra of a few
typical pixels displayed in Fig. 5 (P\'erault et al. in preparation). On the other hand, such bands are
normally completely absent in the LW9 filter (14-16$\mu$m), while the
only PAH band contributing to the LW3 emission is the 12.7~$\mu$m band
at the blue side of the LW3 filter. It is known (Boulanger \&
P\'erault 1988) that the diffuse emission in the 12--20 $\mu$m range
is generally dominated by very small interstellar grains. However, in HII regions the ionic line emission 
(12.8 $\mu$m Ne II, 15.6 $\mu$m Ne III) can also make an appreciable
contribution to the LW3 or LW9 emission.

Fig. 2 displays  the LW2/LW3 images of a ``typical'' ISOGAL field
showing a general similarity in the morphology of the 7 and 15~$\mu$m
emission with a few infrared dark clouds (Sect. 5.2, see also Fig. 5). Figures 3 and 4 show two special cases: 
1) two images, respectively in 
LW6 and LW9 filters, of the star-forming region M16, the Eagle Nebula, where the
LW6 filter, centered on the main PAH feature, enhances the PAH
emission and shows its extraordinary filamentary structure, while the
emission in the narrow LW9 filter, which includes the 15.6 $\mu$m Ne
III line, displays a completely different structure centered on the H
II region; 2) a mosaic of ISOGAL images (assembled by S. Ganesh, in
preparation) in the neighbourhood of the Galactic Center in the
narrow filter LW5 which, unlike LW6, minimizes the PAH
emission. The spectacular dark condensations, especially
the string between the Galactic Center and Sgr B2, strikingly
coincide with mm/submm/far-IR emission features (Lis et al. 1994, 2001, Carey et al. 2000, 
Pierce--Price et al. 2000, Ganesh et al. in preparation).

\subsection{Dark clouds and millimetre follow-up}

The ISOGAL survey revealed a population of narrow (down to 10$\arcsec$
in size), very dark, filaments and globules, seen in absorption in front
of diffuse Galactic emission at 7 and 15~$\mu$m, which are visible in
practically all ISOGAL images (Figs. 2, 3, 5). They were first reported by P\'erault
et al. (1996), who showed that these features are very compact cores
lying mainly between 2 and 8 kpc from the Sun. They are similar to
those (more than 2000) visible at lower angular resolution in the MSX
survey (Egan et al. 1998, Carey et al. 1998, Simon et al. 2001 and references therein). Hennebelle et al. (2001) performed
multiresolution extraction leading to a systematic analysis of more
than 450 such ISOGAL dark clouds located in the inner Galaxy, mostly
in quiescent molecular complexes in the molecular ring and in
 regions interior to that. These authors derived opacities (Sect. 5.3) at 15
$\mu$m in the range of $\tau$ $\approx$ 1--4 for a few selected
objects, leading to column densities of the order of 10$^{23}$
cm$^{-2}$. For a typical dimension of a fraction of a parsec, such dark
condensations have typical densities above 10$^5$ cm$^{-3}$ and masses
more than 10$^3$ M$_\odot$. This interpretation was confirmed by
follow-up observations at radio--mm wavelength carried out at the
IRAM 30~m telescope by Teyssier et al. (2002). They derived physical
conditions and chemical properties of the gas associated with some of
these dense dust condensations, utilising maps of
$^{13}$CO, C$^{18}$O and 1.2~mm continuum emission: these maps show
spectacular correlations with the mid--IR absorption. The molecular
lines are surprisingly weak, indicating likely depletion onto cold
grains. The kinetic temperatures are found to be between 8 and 25~K, with
warmer clouds being associated with young embedded stars.

\subsection{Extinction}

Because of reduced extinction, infrared wavelengths are the only practical way of mapping regions of high extinction in the inner Galaxy. However, in certain regions having 
large values of A$_V$ ($\sim$10--50), the effects of extinction are
still important at near-IR wavelengths, and are not negligible at mid-IR
ones. Extinction corrections must, therefore, be considered in the determination of
intrinsic IR magnitudes and of bolometric magnitudes. The IR reddening or IR counts can be used to determine extinction in visual or infrared. In addition,
when the distribution of A$_V$ along a line of sight is known, its
determination for a specific star can provide a crude estimate of 
distance and, hence, of the stellar luminosity. An early map of A$_V$ in two 
square degrees around the Galactic Center was derived by  Catchpole et al. (1990)
(see also Glass et al. 1987), from near-IR reddening of red giants whose
magnitudes correspond to AGB stars just above the RGB tip. Arendt et al. (1994) used a similar procedure, that is to assume the (J--K)$_0$ colour was constant, to estimate the extinction
towards the Bulge from the low 
resolution ($\sim$0.7 degree) J and K  COBE/DIRBE observations. Cambresy (1999) interpreted extant source counts
from near-IR surveys to map extinction in nearby molecular clouds.

%

        \begin{figure}[]
     \centerline{\psfig{figure={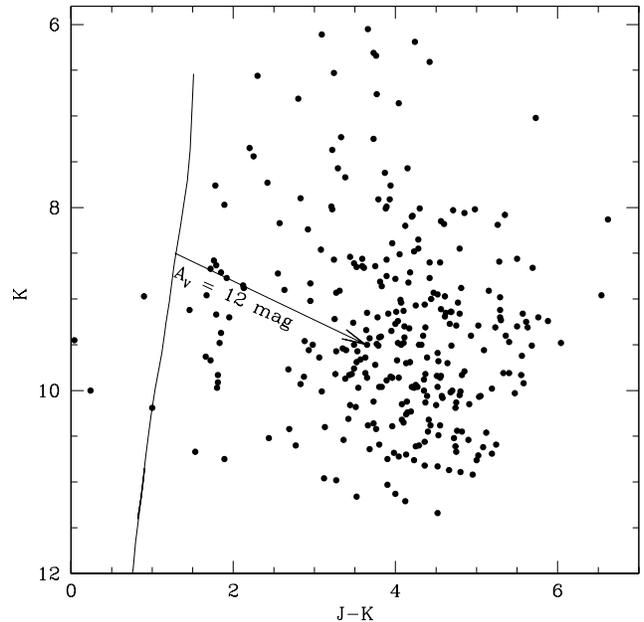},width=9cm}}
        \caption{Colour magnitude diagram (J-K$_{\rm s}$) / K$_{\rm s}$ of ISOGAL sources for a typical field, FC+00733+00015 (see Fig. 7). An isochrone (Bertelli et al. (1994), placed at 8\,kpc
distance, is shown for a 10~Gyr population with Z=0.02. It may be
used to determine the value of A$_V$ on the line of sight of any star
at the Bulge distance, as shown in the example}
	\label{figure6}
        \end{figure}

We have used the DENIS JK$_{\rm s}$ data to extend the method of Glass et al. (1987),
to map extinction in the most obscured part of the Bulge ($\sim$30
deg$^2$; Schultheis et al. 1999). The method is based on the very well
defined K$_0$/J$_0$--K$_0$ relation for most Bulge red giants and AGB
stars (zero-extinction curve in Fig. 6). The JK interstellar
reddening, and hence the A$_V$, can thus be deduced immediately from the J and K$_{\rm s}$ of
each star (Fig. 6). Along each line of sight, the peak of the
distribution in J-K$_{\rm s}$ is interpreted as the Bulge value, giving the
value of A$_V$ to the Bulge in this direction. Of course, this method
is only approximate because of statistical fluctuations and of the
difficulty of disentangling Bulge and Disk distributions at large
$\ell$. In addition, the method is not applicable to the largest
extinctions (A$_V$ $\ge$ 25) where the DENIS J detections are not
complete. Under these conditions, H--K$_{\rm s}$ 2MASS values may be useful with some caution to complement the J--K$_{\rm s}$ data (see e.g. Alard 2001).

There are still uncertainties in the infrared interstellar extinction
law. In the near-infrared, proposed values for A$_{\rm K}$/A$_V$ differ by
as much as 30 \% (see e.g. Glass 1999). However, the difference is not
larger than 7\% for the J K$_{\rm s}$ reddening ( A$_{\rm J}$--A$_{\rm K}$)/A$_V$. The
situation is even more uncertain for the mid-IR ISOGAL bands. In
particular, Lutz et al. (1996; see also Moneti et al. 2001) have
provided, for the Galactic Center line of sight, much larger values
for the extinction in the 7~$\mu$m region than those expected from the
standard interstellar extinction curve 
 for the Solar neighbourhood (see e.g. Draine \& Lee 1984, Mathis 1990).

ISOGAL results provide a way to check the extinction value, averaged
over the broad band used, on the line of sight of many fields along the Galactic Disk, and to correlate it with the distribution of the interstellar gas inferred from CO and H{\small I} surveys (Jiang et al. 2003). The
most straightforward technique is to measure the ratio 
(A$_{K s}$--A$_7$)/(A$_{\rm J}$--A$_{\rm Ks}$),
assuming a constant value for J$_0$--K$_{s0}$ for the intermediate AGB
stars and the luminous RGB stars. In
J--K$_{\rm s}$/K$_{\rm s}$[--7] diagrams, most sources more or less follow a
straight line with some dispersion. For instance, in a well behaved
case ($\ell$ = $-$18.63$^\circ$, {\it b} = 0.35$^\circ$) studied by Jiang et
al. (2003), the slope (A$_{K s}$--$\bar{A_7}$)/(A$_{\rm J}$--A$_{\rm Ks}$)
$\sim$ 0.37 yields (A$_{\rm Ks}$--$\bar{A_7}$) $\approx$~0.06~A$_V$ for the average
extinction $\bar{A_7}$ in the band of the LW2 ISOCAM filter (5--8.5
$\mu$m). Such a value is more compatible with the results of Lutz et
al. (1996) than with the classical extinction curve.  From the value
of A$_{\rm K}$/A$_V$ = 0.089 recommended by Glass (1999) in agreement with
van de Hulst (1946) one deduces $\bar{A_7 }$/A$_V$ $\sim$0.027, while
the value of Rieke \& Lebobvsky (1985), A$_{\rm K}$/A$_V$ = 0.117, yields
$\bar{A_7}$/A$_V$ $\sim$0.056. We favour the Glass--van de Hulst value
for A$_{K}$, but stress that the resulting value, $\bar{A_7}$/A$_V$
$\sim$0.03, is not significantly in disagreement with either the
classical curve or Lutz et al.'s results. In other ISOGAL fields the
J--K$_{\rm s}$/K$ _s$--[7] diagrams often tend to display a large
dispersion in the values of K$_{\rm s}$--[7], which makes the value of
$\bar{A_7}$/A$_V$ less accurate. Nevertheless, the results are consistent with the value derived above (Jiang et al. in preparation). An estimate of $\bar{A}$$_{15}$/
A$_V$ from J--K$_{\rm s}$/K$_{\rm s}$--[15] diagrams by the same method is
practically impossible because of the very large and various values of
(K$_{\rm s}$--[15])$_0$ for many stars (see Sect. 7). 

An independent analysis of the constraints on the interstellar
extinction curve provided by the ISOGAL infrared dark clouds has been
performed by Hennebelle et al. (2001). They show that the
 ratio between the dimming of the diffuse emission at 7 and 15~$\mu$m is 0.75$\pm$0.15 for the clouds located away from the
Galactic Center ($\vert \ell \vert > 1^\circ$) and 1.05$\pm$0.15 for
the clouds closest to the Galactic Center. They derive that the 7~$\mu$m
to 15~$\mu$m opacity ratio is equal to 0.7~$\pm$0.1 for the clouds located
away from the Galactic Center. They discussed several explanations for
the variation of the contrast ratio, including absorption along the
line of sight and local variations of the extinction curve.

\section{Stellar Populations}

\subsection{Colour--Magnitude Diagrams}

        \begin{figure*}[ht!]
        \centerline{\psfig{figure={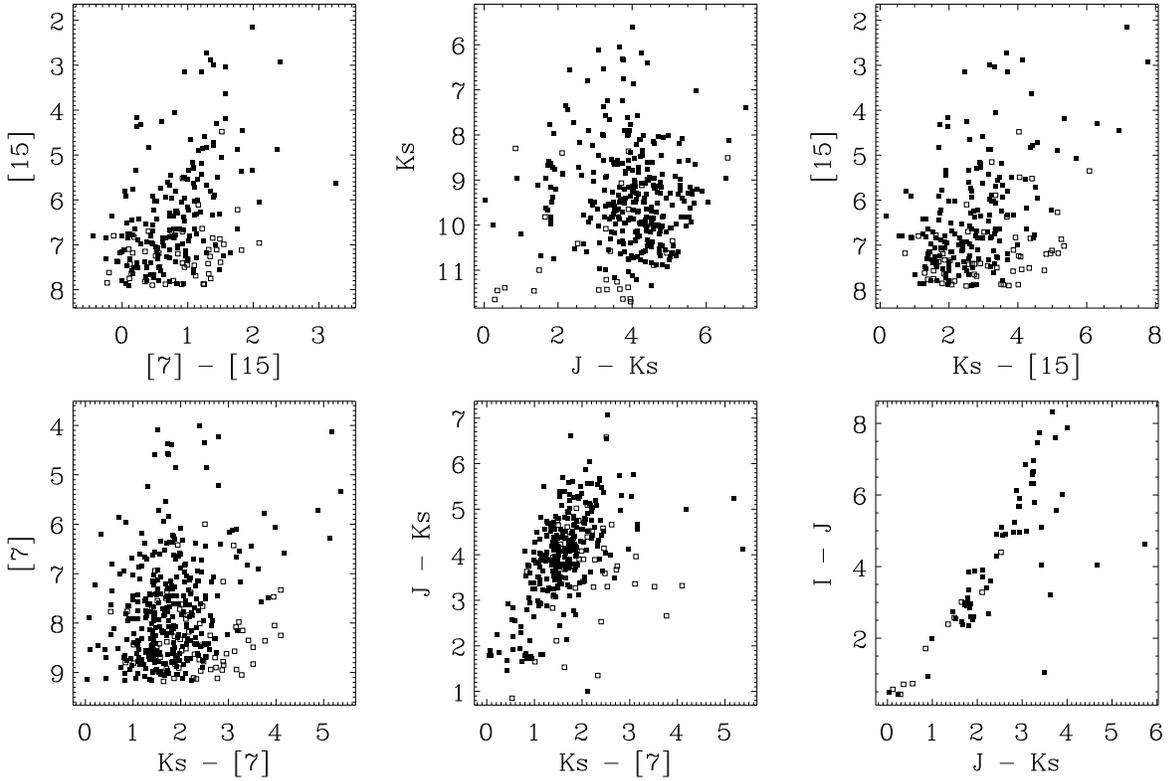},width=16cm}}
        \caption{Most useful ISOGAL--DENIS colour-magnitude and colour-colour diagrams for a
typical field, FC+00733+00015 (see Fig. 2). Sources of reasonably good quality in
all points (7 \& 15 $\mu$m photometry and reliability, 7--15 $\mu$m
associations, ISOGAL--DENIS associations) are shown as black
points. Sources of poorer quality are displayed as open squares. 
The luminosity of the Bulge RGB tip  corresponds to K$_{\rm s}$~$\approx$~10 and [7]~$\approx$~[15]~$\approx$~8}
        \label{figure7}
        \end{figure*}

The availability of multi-wavelength information allows one to make the distinction
between intrinsic source properties and effects due to interstellar
extinction.  For this purpose the several colour--magnitude diagrams
(CMD) and colour--colour diagrams that one can construct with the
diverse combinations of the five bands from ISOCAM and DENIS are
effective.  The most useful examples are displayed in Fig. 7. As
discussed in Sect. 5.3, the diagrams K$_{\rm s}$/J--K$_{\rm s}$, [7]/K$_{\rm s}$--[7] or
K$_{\rm s}$/K$_{\rm s}$--[7], J--K$_{\rm s}$/K$_{\rm s}$--[7] (and also I--J/J--K$_{\rm s}$) are useful
for determining interstellar extinction. I--J can provide, in particular,
information about metallicity through the effect of absorption in
molecular bands of TiO, VO, etc. (see van Loon et al. 2003). However, 
it is very difficult to separate the effects of T$_{eff}$ and metallicity. In addition,  
it
should be realised that in many ISOGAL lines of sight, only a fraction
of ISOGAL sources have useful I data, because the detection at I is
limited to moderate values of A$_V$ (typically $\le$~10).

The most important new feature of ISOGAL data compared to
previous near-IR studies is their ability to detect and characterise
sources with a mid-infrared excess arising from even quite small
amounts of circumstellar dust. The presence of such dust, even in very
small amounts, is always revealed clearly  by a 15~$\mu$m excess, i.e.,
namely, by the colours (K$_{\rm s}$--[15])$_0$ and especially [7]--[15] (see
Sect. 7). Indeed, the [7]--[15] colour, being weakly
sensitive to extinction, can provide a direct estimate of mass-loss
rates of AGB stars even without the need for explicit dereddening.

%


        \begin{figure*}[ht!]
        \centerline{
	  \resizebox{15.5cm}{!}{\includegraphics{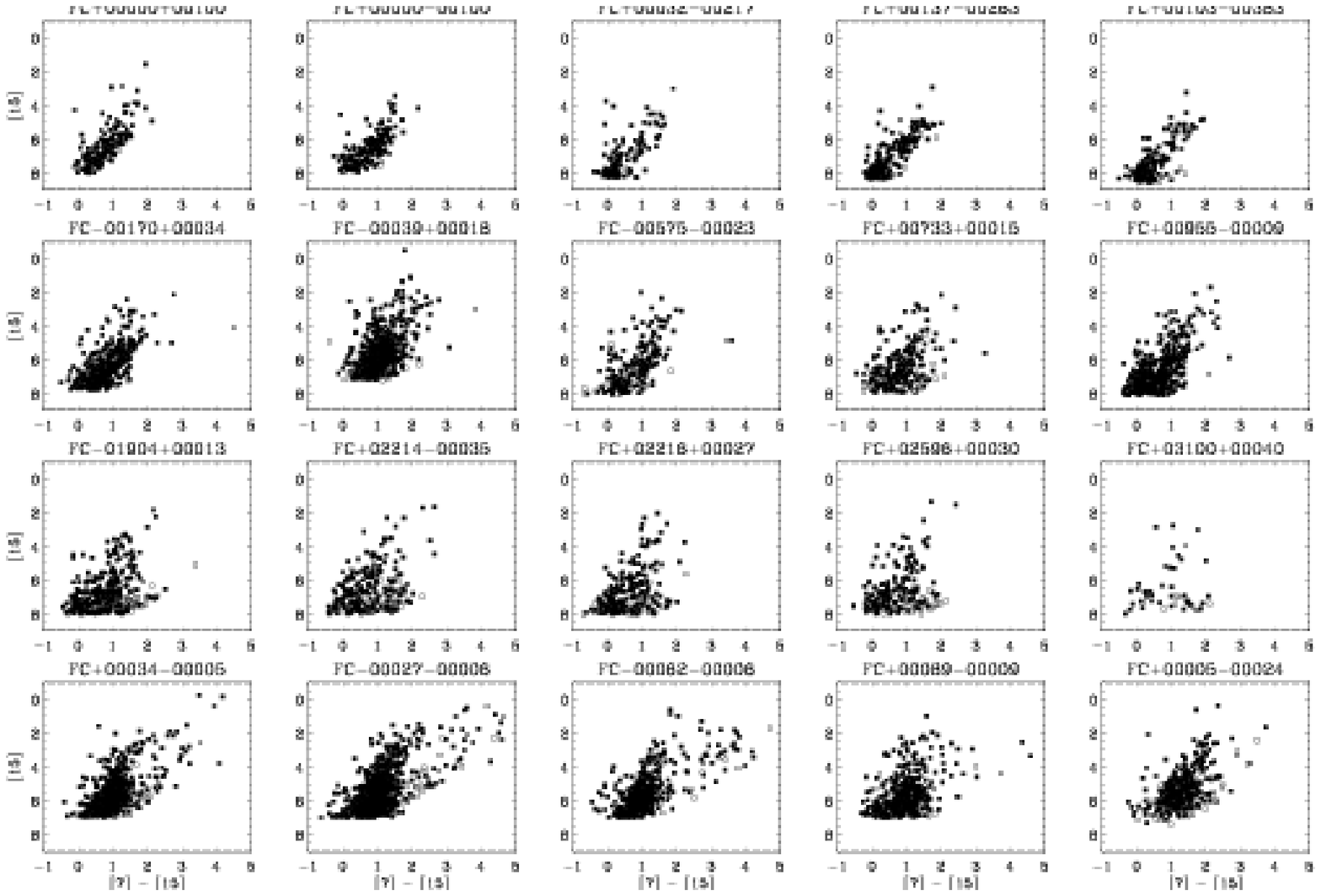}}}
        \caption{Typical colour-magnitude diagrams [15]/[7]--[15]$~$ in various Galactic regions: first row
intermediate Bulge; second and third rows Galactic Disk; fourth row
inner Bulge. 
Sources of reasonably good quality in
7 \& 15 $\mu$m photometry and reliability, as well as 7--15 $\mu$m
associations, are shown as black
points. Sources of poorer quality are displayed as open squares. }
        \label{figure8}
        \end{figure*}

The colour-magnitude diagram [15]/[7]--[15] thus provides a straightforward
view of the types and the number of dusty sources in each ISOGAL
field, with even a direct measure of their mass-loss rates for AGB
stars. Such diagrams are displayed in Fig. 8 for various typical
ISOGAL fields in the Galactic Bulge and Disk. Their most conspicuous
common feature is a sequence of AGB stars, with at least some inferred 
mass-loss; these are the majority of the sources detected at
15~$\mu$m.

The interpretation is especially unambiguous in the intermediate
Bulge, with $\vert$b$\vert$ $\sim$ 1$^\circ$ (first row of Fig. 8$~$;
Glass et al. 1999, Omont et al. 1999, Ojha et al. 2003, van Loon et
al. 2003). There, the sequence is particularly well defined and
narrow, and it includes more than 90\% of the sources. The obvious
interpretation is that most stars are luminous red giants (AGB or RGB)
at approximately the Galactic--Center distance, and that the extinction is
small. The sources at the beginning of the sequence, with practically
no 15~$\mu$m excess, have the luminosity expected for Bulge giants at
the tip of the first ascent red giant branch ([15]~$\approx$~8.5). The 15~$\mu$m excess at the
other end of the sequence is typical of that corresponding to the
mass-loss rate from Mira variables ($~\sim$ a few $10^{-7}
M_{\odot}$/yr). The other stars of the sequence are thus
``intermediate'' AGB stars with very little, but still detectable,
mass-loss. A few foreground or younger stars are also visible in the
diagrams towards bluer colours and brighter magnitudes than the AGB sequence. Since dereddening is
relatively easy in such lines of sight, the diagrams
[15]$_{0}$/(K$_{\rm s}$--[15])$_0$ and K$_{s0}$/(K$_{\rm s}$--[15])$_0$ are
generally similar in their interpretation, tracing the same
populations and their mass-loss and are even more sensitive than the [15]/[7]--[15] diagrams for mass-loss determination (Sect.
7).

The AGB sequence is also conspicuous in [15]/[7]--[15] diagrams in all other
lines of sight (Fig. 8). Towards the very inner Bulge ($\vert \ell
\vert$ $<$ a few degrees, $\vert$ {\it b}$\vert$ $<$ $0.5^\circ$, fourth row 
of Fig. 8), the sequence is vertically broadened by the very large and
variable extinction A$_V$~$\sim$ 20~-~35 mags. Luminous and very cold AGB
stars, mainly OH/IR stars, are also more numerous (Sect. 7.4). In
addition, a population of luminous young stellar objects is
conspicuous to the right of the diagrams (Sect. 8).

A similar AGB sequence is also visible in the lines of sight toward
the Galactic Disk away from the Bulge. 
However, the diagrams are made less readily interpretable  by the
broader range of distances to the sources (see e.g. Jiang et
al. 2003). The fraction of foreground sources (mostly RGB and AGB)
is larger, which broadens the distributions towards the left. Depending
on the line of sight, young stars  may or may not be 
present in the (lower) right part of the diagrams; however, their
distinction from AGB stars is not always obvious at the boundary of
the AGB sequence (Sect. 8).

Although the information provided by the five--wavelength ISOGAL data is especially useful to estimate A$_V$ and the nature of the sources, it cannot remove all the ambiguities, in particular because of the non-simultaneity of the observations at various wavelengths. Complementary spectral information is necessary to check the photometric analysis and to address further questions such as metallicity. For most ISOGAL sources with A$_V$~$\ga$~10, one needs  near-IR spectroscopy. The recent results of Schultheis et al (2003) on lines of sight close to the Galactic Center illustrate the importance of such systematic follow-up programs: identification of mass-losing supergiants, of YSOs, of AGB stars with extremely red ISOGAL colours and the difficulty of separating YSOs and such AGB stars, metallicity estimates of inner Bulge AGB stars.

\subsection{Red Giants}

Sources on the upper first ascent red giant branch (RGB) are the most
numerous at 7~$\mu$m. Since there is no easy  way to distinguish them,
among ISOGAL sources, from the less numerous AGB stars of similar
luminosity below the RGB tip based on ISOGAL measurements, we simply call stars of both types below
the RGB tip ``red giants''. (Symmetrically, ``AGB stars'' denotes
stars on the asymptotic giant branch above the RGB tip). Since RGB red
giants are very numerous in the ISOGAL data, and form a relatively
homogeneous class of objects whose properties show less intrinsic
dispersion than is seen on the AGB, 7~$\mu$m sources are the best
tracers of Galactic structure and of the old dominant stellar
population. The low extinction at 7~$\mu$m makes such
tracers attractive in highly obscured regions of the inner Galactic Disk and
Bulge. However, their interpretation is not really straightforward and
needs some elaboration for two reasons: the extinction correction,
although rather small, is not negligible, typically $\sim$0.3--0.5
mag (see Sect. 5.3); the uncertainty on the distance remains a
major problem. Distance uncertainties are large and the reddening information
yields only a rough distance determination when correlated with known
structures. However, close to the Galactic Center direction, the
highly peaked density in the central Bulge dominates the foreground
and the background structures to such an extent that it allows, with
some care, a detailed analysis of inner structure (see e.g. Alard
2001 for a near-infrared analysis). On the other hand, in the Disk
away from the Bulge the distance uncertainty prevents any simple
discussion of stellar ages based on AGB luminosities, and even a
clear distinction between red giants and AGB stars.

An additional difficulty is that, especially in the
Bulge region, the RGB tip magnitude, with [7]~$\sim$8--8.5, is not very far from
the ISOGAL sensitivity limit (Sect. 2) so that a completeness
correction has to be applied, adding to the total uncertainty. The
 7~$\mu$m magnitude range usable for tracing structure  will  thus be reduced to about one
magnitude. The surface density of red giants in the correct magnitude range is typically a few 10$^3$
deg$^{-2}$, which gives good statistics in cells of a few 10$^{-2}$
deg$^2$, i.e. the typical dimension of the smallest ISOGAL fields. The 
latter give a sparse sampling of the stellar density of the inner Disk 
and Bulge, which can be compared with the density inferred from
near-infrared surveys (see Alard 2001). Anyway, the best strategy is to
combine near-IR and 7~$\mu$m data: the 2MASS or DENIS K$_{\rm s}$ data allow one to
check the completeness of the 7~$\mu$m data and to correct 7~$\mu$m
counts; the 7~$\mu$m data give a check on the extinction correction for
the near-IR data.

Because of the difficulties described above, a comprehensive analysis
of Galactic structure from ISOGAL data had to wait for a homogeneous
and certified catalogue of ISOGAL sources. It is thus still in
progress (see van Loon et al. 2003). 
Extension of current  analyses of the red giant stellar
populations at the scale of the ISOGAL fields, in the obscured inner
Bulge and Disk, will be feasible with deeper near-IR and mid-IR observations
with smaller pixels and larger cameras, with large ground telescopes in the 
near-IR and new space missions such as SIRTF or ASTRO--F.

\subsection{Inner Bulge AGB: luminosity, photospheric colours and age}
 
The AGB luminosity function is a very sensitive diagnostic of the age
distribution of a stellar population. For ISOGAL, if the distance and the extinction are known,
the luminosity is directly derived from the flux density in the five
DENIS--ISOCAM bands, which cover the relevant spectral energy
distribution (see e.g. Loup et al. 2003). Straightforward methods of interpolation and
extrapolation provide a very good estimate of the total flux. Indeed, except for the extremely red 
ISOGAL sources, the dereddened
K$_{s0}$ magnitude produces a good value of the total flux,
since the K$_{\rm s}$ bolometric correction equals $\sim$3 for most sources with
little scatter (see e.g. the discussion in Ojha et
al. (2003) and the references therein). For a distance modulus of
14.5, corresponding to d$_{GC}$ = 8 kpc, the absolute bolometric magnitude of
most ISOGAL-DENIS sources is thus approximately given by
\begin{equation}
	{\rm M_{bol} = K_{s0}  - 11.5}
\end{equation}
Some care is required before determining the luminosity function of
the Bulge ISOGAL sources. One needs to properly identify and discard
foreground (and background) sources, to determine the interstellar
extinction in the line of sight towards  each source, to estimate the
effects of the photometry and variability uncertainties, and to correct for count
incompleteness.

The most obvious foreground sources are readily identified by
inspection of the distribution of J--K$_{\rm s}$ for a given ISOGAL field or
part of a field. In lines of sight towards the Bulge, the distribution of source
counts is strongly peaked around a value (J--K$_{\rm s}$)$_{_rm B}$ which
corresponds to the mean extinction up to the central Bulge
(Fig. 6). A minority of the sources lie below the bulk of
J--K$_{\rm s}$ values, because they are in front of
part of the dust down the line of sight towards the Bulge (see Sect.
5.3). However, this method is increasingly unreliable for more distant
sources: depending on the inhomogeneity of the dust distribution in
the Disk, Disk sources, nearer than the Bulge but close to it, cannot
be distinguished from Bulge sources purely on the value of the
extinction. The situation is similar for stars behind the Bulge (see
e.g. Ojha at al. 2003).

The extinction in the line of sight towards a given star may often be
accurately derived from the J and K$_{\rm s}$ magnitudes, either simply  from
the value 
\begin{equation} 
{\rm J - K_{s} = J_{0} - K_{s0} + A_{J} - A_{Ks}}
\end{equation} assuming an estimated value of J$_0$--K$_{s0}$, or,
better, from an isochrone K$_{\rm s}$/J--K$_{\rm s}$ (see e.g. Schultheis  et
al. 1999, van Loon et al. 2003). However, this method fails when the
extinction is so large (A$_V$ $\ga$ 25) that the star is not detected
in J by DENIS or 2MASS (Sect. 5.3). One could then use another
colour such as H--K$_{\rm s}$ (Alard 2001) or K$_{\rm s}$--[7] (van Loon et
al. 2003, Jiang et al. 2003). However, the method fails for very red
AGB stars when the intrinsic colours such as J$_{0}$ - K$_{s0}$ or
K$_{s0}$--[7]$_0$ are affected by the presence of mass-loss. A
simpler, necessarily less accurate, method (Wood et al. 1998) is then
to assume that such red AGB stars have the same interstellar
extinction as the average bluer DENIS stars, such as determined 
e.g. by the general DENIS extinction map of the Bulge (Schultheis et
al. 1999).

The present infrared surveys will significantly improve knowledge of
the inner Bulge AGB population and allow the use of AGB stars as
tracers to infer the age distribution of the inner Bulge stellar
population (van Loon et al. 2003). Previous studies were mostly
limited to the outer-Bulge low extinction windows (Frogel \& Whitford 1987, Frogel et al. 1990, 1999, Tiede et al. 1995)
or to very deep near-IR studies in very small fields (Rich et
al. 1996), except for a few wide-area near-IR observations around the
Galactic Center (Glass et al. 1987, Catchpole et al. 1990, Phillip et al. 1999). 

Complementary data on Bulge AGB stars, allowing one to disentangle
mixed populations, is becoming available from velocity information, extending previous work such as Sharples et al. (1990).
Multi--object infrared or visible spectroscopy in regions where the
extinction is not too high is currently feasible.  Radial velocities of
several hundreds of ISOGAL AGB stars are already measured from the
detection of OH or SiO maser emissions (Sect. 7.4).

\section{AGB circumstellar dust and mass--loss}

As discussed in Sect. 6.1, the excess of emission at 15~$\mu$m shows
that the majority of sources detected at 15~$\mu$m by ISOGAL are
mass--losing AGB stars. ISOGAL thus provides a characterisation of AGB
mass--loss in the whole inner Galaxy and especially in the inner
Bulge. Such a very large sample, of several times 10$^4$ sources, with many
extreme stars, will allow a large variety of follow-up studies of
circumstellar envelopes to analyse their properties through the
Galaxy. In particular, the identification of the exact nature of
circumstellar dust in the bulk of AGB stars and in peculiar sources,
as well as a good calibration of mass--loss, requires a comprehensive
follow-up program of infrared spectroscopy: this is possible with SIRTF or
ASTRO-F. ISOGAL CVF spectroscopy in a few small fields already brings
preliminary information on AGB circumstellar dust in the central
Galaxy.

\subsection{CVF spectroscopy}	

In addition to the large imaging survey, ISOGAL has a small
complementary spectroscopy program with the CVF mode of ISOCAM. It has
provided 2-D integral-field spectroscopy (6$\arcsec$ pixels, 5-16$\mu$m,
$\lambda/ \Delta \lambda$~$>$~35) in eighteen ~3'~x~3' sub-fields
(32~x~32 pixels each) in chosen ISOGAL fields. This
program had as dual goal of characterisation of the mid--infrared
emission of the diffuse interstellar medium (dominated by PAHs, see
Sect. 5.1 and Fig. 5) and of the strong point sources (AGB
stars). Therefore, the CVF subfields were selected partly in regions
of strong diffuse emission, and partly in fields of the intermediate
Bulge with low diffuse emission and a high density of late type
stars. It is known that the analysis of such CVF observations is
particularly difficult in regions with strong diffuse emission and
bright sources, such as ISOGAL fields, because of residual stray
light. Consequently, the analysis of ISOGAL CVF spectra is still in
progress (Blommaert et al. in preparation, P\'erault et al. in preparation). Preliminary results have been
discussed by Blommaert et al. (2000). The analysis of one 3'~x~3' sub-field specially selected for the detection of Bulge late-type stars, has shown that around 30 objects were detected of which half had a significant flux density up to 16$\mu$m. One may expect more than one hundred spectra of late-type stars from the other subfields. Most sources showing a mid--IR excess do not show the typical silicate feature often observed in Mira--type variables. The feature is much broader and peaks at longer wavelength ($\sim$12$\mu$m) than the traditional 9.7~$\mu$m peak. Such broad emission features are common in IRAS data (Sloan \& Price 1995 and references therein). They have been associated with aluminium oxide grains  (Onaka et al. 1989, Egan \& Sloan 2001 and references therein). It is predominantly observed in CVF spectra of sources with low mass-loss, in agreement with the interpretation that ISOGAL is mainly detecting the onset of the mass-loss on the AGB (Sect. 7.3).

        \begin{figure}[]
        \centerline{\psfig{figure={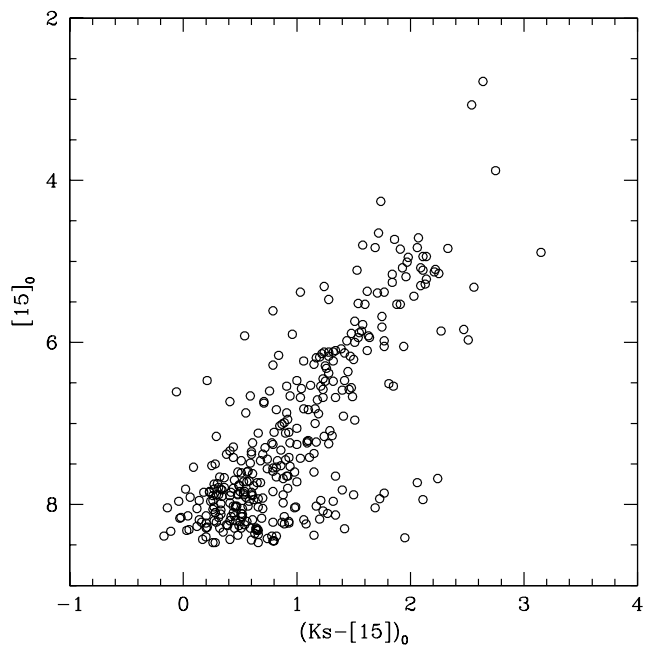},width=9cm}}
                \caption{Dereddened colour--magnitude diagram combining the two ISOGAL
fields observed in the Baade's Windows (Glass et al. 1999, Ojha et
al. 2003) showing the distribution of (K$_{\rm s}$--[15])$_0$. The
uncertainty introduced by dereddening is negligible and
(K$_{\rm s}$--[15])$_0$~$>$~1 is a signpost for circumstellar dust. The points with the largest values of (K$_{\rm s}$--[15])$_0$ at the 
right-bottom of the diagram may be the results of wrong associations.  The RGB tip is just at the bottom of the sequence at [15]$_0$~$\approx$~8.5}
        \label{figure9}
        \end{figure}

\begin{figure*}[htbp]
\begin{center}
\resizebox{15cm}{!}{ \rotatebox{0}{\includegraphics{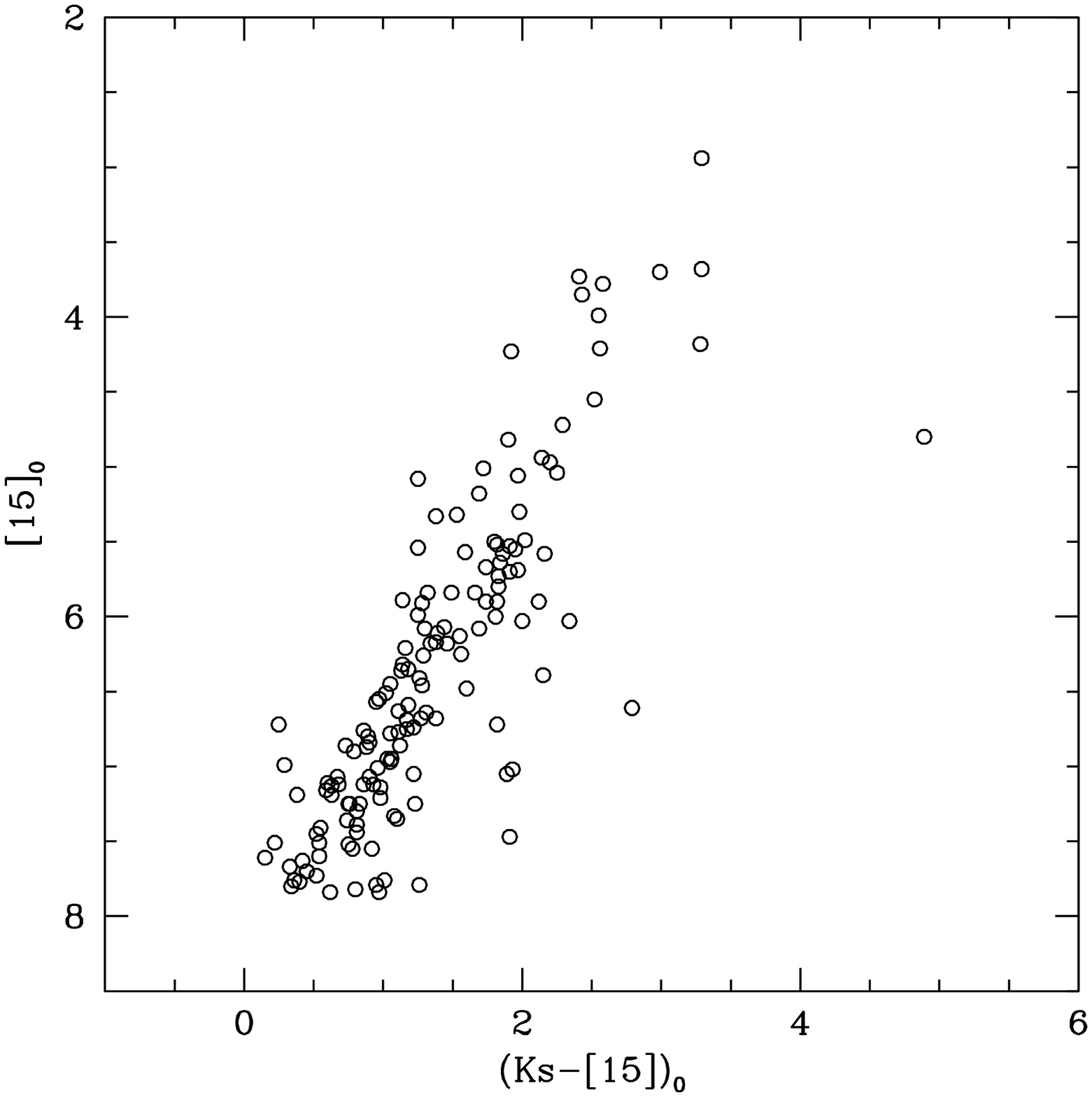}}
		\hspace{10mm}
 		\rotatebox{0}{\includegraphics{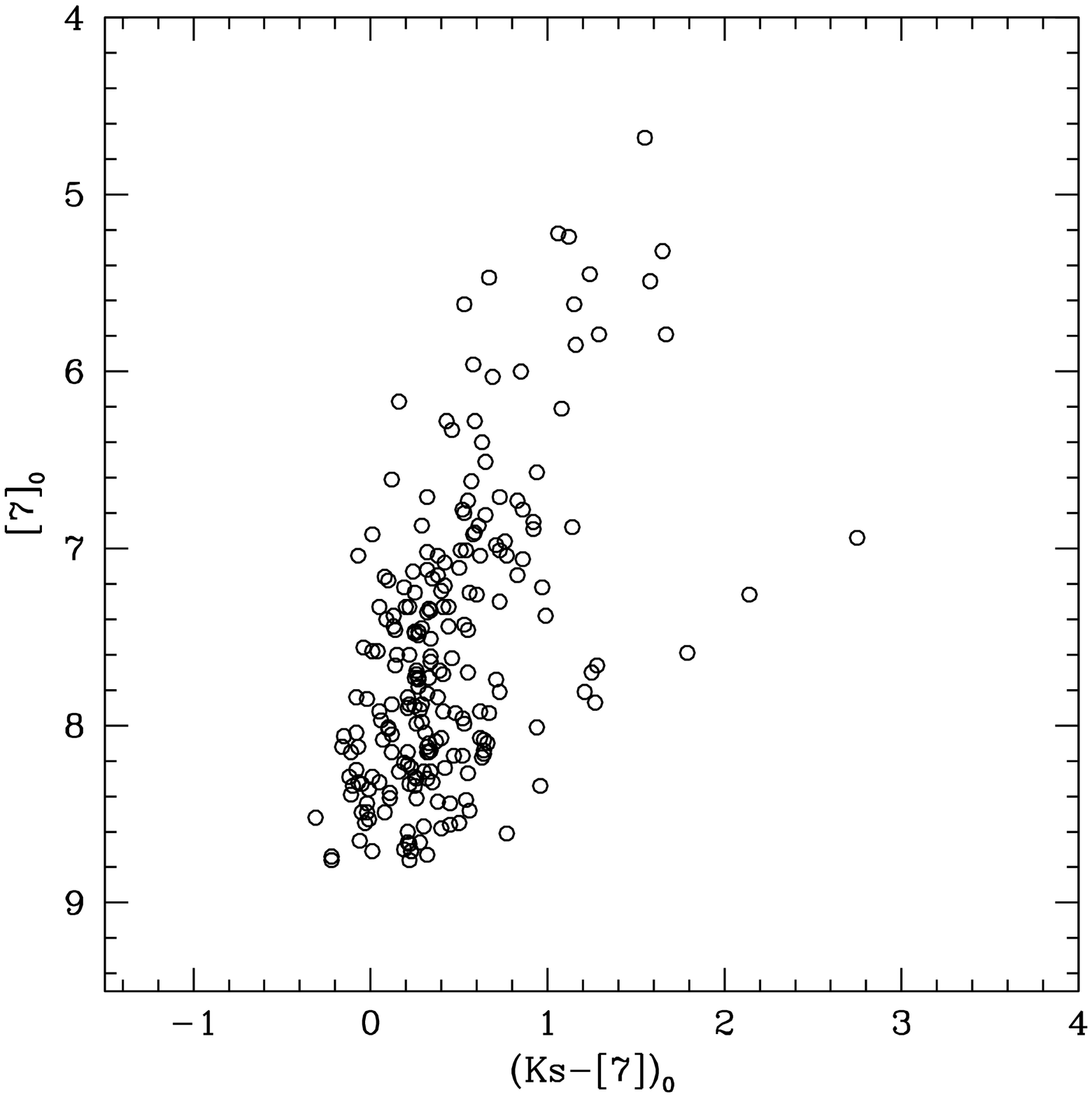}}}
\caption{Comparison of the dereddened colour--magnitude diagrams
[15]$_0$/(K$_{\rm s}$--[15])$_0$ and [7]$_0$/(K$_{\rm s}$--[7])$_0$ in the Bulge
field FC+00000+00100 with mild extinction (Omont et al. 1999, Ojha et
al. 2003). The colour (K$_{\rm s}$--[15])$_0$ is much more sensitive to the
presence of circumstellar dust than (K$_{\rm s}$--[7])$_0$. The points with the largest values of (K$_{\rm s}$--[7])$_0$ at the 
right-bottom of the right diagram may be the results of wrong associations.  The RGB tip is just at the bottom of the sequence of the right panel at [7]$_0$~$\approx$~8--8.5 }
\end{center}
\label{figure10}
\end{figure*}

\subsection{15~$\mu$m excess}

The very red colours (K$_{\rm s}$--[15])$_0$ and [7]--[15] cannot be explained
just by the emission of very cold photospheres, but are clearly
dominated by dust emission. This is well demonstrated, e.g., in Fig. 9
for the two observed ISOGAL fields in the Baade's Windows (Glass et
al. 1999). For a number of stars with known spectral types, such
observed infrared colours have been compared with the expected photospheric
colours and the model isochrones derived from Bertelli et al. (1994)
(van Loon et al. 2003). The 15~$\mu$m excess is obvious for a large
fraction of the sources. It increases with the luminosity and for
the latest spectral types, as expected from the properties of
mass--losing AGB stars in the solar neighbourhood. Similar 15~$\mu$m
excess measurements, showing the presence of circumstellar dust, are
obvious in the  analysis of every ISOGAL field (see Fig. 10, Omont et
al. 1999, Alard et al. 2001, Ojha et al. 2003, Jiang et al. 2003). On the other hand, a similar excess at 7~$\mu$m is much less
obvious for most sources of Fig. 10 right, which are still dominated by
photospheric emission at 7~$\mu$m, except for the most luminous and
coldest cases.

\subsection{Mass--loss rates}

Knowledge of the mid-- or far--infrared emission characteristics of
dust in circumstellar envelopes AGB stars is one of the best ways for 
estimating their mass--loss rates (see e.g. Jura 1987, Habing 1996, Le
Bertre et al. 2001, Le Bertre \& Winters 1998, Jeong et al. 2002 and references
therein). As discussed in Ojha et al. (2003), the value of (K$_{\rm s}$--[15])$_0$ is the best way for 
deriving mass--loss rates from ISOGAL--DENIS data from the relation between $\dot{M}$ and (K$_{\rm s}$--[15])$_0$ established from recent modelling by Jeong et al. (2002 and in preparation), which is in agreement with 
Whitelock et al. (1994) for (K-[12])$_0$. When K$_{s0}$ is not accurately known, one can  use
[7]--[15], although this is less precise. However, in addition to the errors in (K$_{\rm s}$--[15])$_0$ due to variability and dereddening, such estimates of
$\dot{M}$ directly depend on the assumed model for circumstellar dust
infrared properties and  dust--to--gas
ratio, which remain uncertain despite the important progress achieved by Jeong et al. (2002 and in preparation).
Ojha et al. (2003) have performed a preliminary
estimate of mass-loss rates from (K$_{\rm s}$-[15])$_0$ and Jeong et al's model for the AGB
stars of the ISOGAL fields of the intermediate Bulge ($|$$b$$|$ $\sim$
1$^\circ$ -- 4$^\circ$). Within the old population, stars having small mass-loss rates contribute appreciably to
the  mass returned to the interstellar medium. However, larger mass-loss rates appear to have a dominent contribution for younger populations, and even the few stars (less than one
percent) which have mass-loss rates $\ga$ 10$^{-5}$
M$_{\odot}$/yr, could dominate the mass return to the
interstellar medium if they belong to the inner Bulge, although it is still
unclear if indeed they do.

Systematic derivation of the distribution function of mass-loss rates
for the AGB stars of the central Bulge and Disk, and of their global
mass return to the interstellar medium is thus feasible, and will be a
major outcome of ISOGAL. However, reliable consolidation of quantitative results relies on
accurate observational validation of dust models, dereddening and Disk/Bulge discrimination, which are not simple tasks, but will certainly progress with future mid-infrared studies.

\subsection{Specific AGB classes : Long-Period Variables, Masers}

The largest AGB mass-loss rates ($>$~10$^{-7}$ M$_\odot$/yr) are associated with large amplitude, long period variability. In a very few ISOGAL fields of the Bulge,
Mira-type long-period variables (LPV) had been previously identified
by Lloyd Evans (1976), Glass \& Feast (1982) and Glass et
al. (1995, 2001). As expected, they correspond to bright ISOGAL sources with
large values of (K$_{\rm s}$--[15])$_0$ ($\sim$1.5--2.0) and [7]--[15]
($\sim$1.0--1.5) (Glass et al. 1999, Schuller et al. in
preparation). Information on large-amplitude variability may also be
obtained in the few fields with repeated ISOGAL observations with identical  parameters 
at different dates (Schultheis et al. 2000, Schuller et al. in
preparation).

Discussions of the variability of ISOGAL sources have been
considerably extended by the analysis by Glass \& Alves (2000, Alard et
al. 2001) of the MACHO data in two ISOGAL fields in Baade's
Windows. MACHO V and R light curves were derived for more than 300
ISOGAL stars, practically all those detected both at 7 and 15~$\mu$m
in the regions of the two fields observed by MACHO. These stars,
mainly on the AGB and at the RGB-tip, are found to possess a wide and
continuous distribution of pulsation periods. About 5\% are Miras and
nearly all the others are semi-regular variables with periods from 10
to 230 days and small amplitudes. Their mass-loss rates depend on
luminosity and pulsation period. Some stars lose mass as rapidly as
short-period Mira variables but do not show Mira-like amplitudes. It
is found that a period of 70 days or longer is a necessary but not
sufficient condition for mass-loss to occur. The DENIS and 2MASS
near-infrared properties of these ISOGAL LPVs and their optical spectral types have been discussed by
Schultheis \& Glass (2001) and Glass \& Schultheis (2002).

OH/IR stars are known to have the most extreme mass-loss rates among
AGB stars, even larger than Miras. Their census is reasonably deep 
in a region $\sim$1$^\circ$ around the Galactic Center, which has
been well covered by ISOGAL observations. Ortiz et al. (2002) have
found that all the known OH/IR stars observed by ISOGAL are among the brightest 15~$\mu$m sources detected. A  mass-loss rate can be 
estimated from (K$_{\rm s}$-[15])$_0$ or ([7]-[15])$_0$, and a bolometric
luminosity L determined from dereddened magnitudes in the
DENIS--ISOCAM bands. L is directly related to the initial mass M$_i$ of
the star and hence to its age. The number of the most luminous OH/IR
stars thus reflects their formation rate in the inner Bulge
$\sim$~10$^8$--10$^9$ yr ago. Preliminary results by Ortiz et
al. (2002) confirm that this number is relatively small.

A systematic search for 86 GHz SiO maser emission has begun 
using the IRAM 30~m telescope  (Messineo et al. 2002).
SiO masers have been
detected in a large percentage of ISOGAL sources with infrared colours
and magnitudes characteristic of Long Period Variables. More than 270
detections have already been achieved and their line--of--sight
velocities indicate that the stars are located in the inner
Galaxy. These new detections double the number of line--of--sight
velocities derived from previous observations of OH and SiO masers in
the central parts of the Galaxy and will facilitate dynamical studies. The newly-obtained longitude-velocity diagram clearly reveals a stellar nuclear Disk component. Follow-up near-infrared spectroscopy of those high-velocity SiO masing stars has been started (Messineo et al. in preparation). The analysis of  the infrared properties of OH and SiO stellar maser emitters will also give new clues on the occurrence of the different masers with AGB type.

\subsection{Related cases: M Supergiants, Planetary Nebulae and PPNe}

It is clear that ISOGAL has detected most M supergiants present
in the observed fields. The census of ISOGAL supergiants is triply 
interesting: it gives direct information on the star formation in the last few
tens of millions of years in the inner Galaxy, especially in the broad
neighbourhood of the Galactic Center; it offers specific information on the
mass-loss from inner-galaxy supergiants; and it may yield the discovery of new exceptional
objects, if any, with very large mass-loss and luminosity similar to VY
CMa or NML Cyg (as, e.g., in the LMC for the exploitation of the MSX/2MASS data by Egan et al. 2001). It is clearly difficult to distinguish supergiants from AGB red
giants from their  luminosity unless their 
distance is known. 

A census of the most luminous infrared sources in the broad vicinity
of the Galactic Center is thus in progress from ISOGAL data complemented by
the MSX survey (Schuller 2002, Chitre et al. in preparation). Preliminary results from Schuller (2002) have identified a couple of hundred M supergiant candidates. As for AGB stars
(Sect. 7.3), reasonable estimates of the mass-loss rates can be
provided by the values of (K$_{\rm s}$--[15])$_0$ or [7]--[15]. The mass-loss
rates of red supergiants may span a large range of values (Jura 1987,
Omont et al. 1993
). Most of supergiant candidates identified by Schuller (2002) have relatively low luminosity in the range $\sim$2--5~10$^{4}$ L$_\odot$ with low mass-loss. A few luminous ($\sim$ 10$^{5}$
L$_\odot$) supergiants with quite appreciable mass-loss rates
($\sim$10$^{-6}$-10$^{-5}$ M$_\odot$/yr) have also been identified in the central Bulge/Disk (Schultheis et al. 2003, Schuller 2002),
outside the three well known clusters of massive young stars
which could not be observed by ISOGAL because of the need to avoid detector saturation. However, no new
extraordinary object with huge mass-loss, comparable to the few known
in the Disk, has been identified in ISOGAL or MSX data within several
degrees of the Galactic Center. In some cases, it may be very
difficult to distinguish a low luminosity supergiant from a high
luminosity AGB star without spectroscopy. Since optical spectroscopy
is practically impossible for sources with very large extinction, it
is necessary to rely on infrared spectroscopy. Preliminary results of the
near-IR spectroscopy program of ISOGAL sources in the central Bulge
(Schultheis et al. 2003) have confirmed the identification
of a few such luminous mass-losing red supergiants.

Planetary nebulae (and post-AGB pre-planetary nebulae, PPNe) are
another class of very red ISOGAL sources not easily distinguishable
from YSOs and extremely red AGB stars, without information at longer
wavelength than 15~$\mu$m. Beyond the few planetary nebulae previously
known in the ISOGAL fields, the search for such objects among ISOGAL
sources thus requires additional information and is just beginning
(Gorny et al. in progress).

\section{Dust enshrouded young stars and star formation}

Young stars with  dusty circumstellar disks or cocoons (YSOs) are the
second large class of Galactic sources identified by their excesses in
the mid- or far-infrared. However, given the emission excess
induced by large amounts of cold dust, relatively far infrared data,
at 25 or 60 $\mu$m, are generally essential to distinguish YSOs from
AGB stars.  The lack of a wavelength longer than 15~$\mu$m in ISOGAL
is a serious handicap to the identification of YSOs from the much more common AGB
stars.  The main criterion available for this purpose with
ISOGAL--DENIS data is a very large excess in the colours [7]--[15] or
(K$_{\rm s}$--15)$_0$ (similarly, very large K$_{\rm s}$ - [8.3] 2MASS--MSX colours have been used by Egan et al. (2001) to identify YSOs in LMC). However, robust discrimination between YSOs and AGB
stars in ISOGAL data remains difficult. Additional criteria which are useful to fully confirm
the identification of YSOs among ISOGAL sources (except for those with
very large 15~$\mu$m excess) include (Schuller 2002): the spatial extent of the ISOGAL
source; association with a known star-forming region; association with
a radio source; far-IR excess in MSX or IRAS data; infrared (or
visible) spectroscopy.

Felli et al. (2000) have stressed that many luminous ISOGAL YSOs are extended sources. However, 
no systematic, complete extraction of extended sources has been performed yet;  but
the present Point Source Catalogue also contains sources of moderate extent,
with sizes of the order of 10-20$\arcsec$ (FWHM). These slightly-extended sources, if sufficiently bright, 
can be separated from the really point-like sources thanks to abnormally
high values of their uncertainty on the measured magnitudes, with typical
$\sigma \approx$ 0.15 mag for bright sources, whereas point sources
in the same brightness range generally have $\sigma < 0.05$ mag. 
Aperture photometry has shown that the magnitude of such sources
can in fact be underestimated by about 1 mag (Schuller et al. 2003, Schuller 2002).

The case of the most luminous, and, hence, most massive, YSOs
[F(15$\mu$m) $\ga$ 0.3 Jy, L $\ga$ 2500 L$_\odot$ at 5 kpc, implying M
$\ge$ 10 M$_\odot$] has been systematically discussed by Felli et
al. (2000, 2002). They have shown, from a radio identified sample of YSOs (Becker et al. 1994, Testi et al. 1999) that the criterion [7]--[15] $>$ 1.8
can be used to identify YSO candidates with a probability greater than
50\%. Furthermore, the more conservative condition
[7]--[15] $>$ 2.5 provides a more reliable discrimination
from AGB (OH/IR) stars (however, see Schultheis et al. 2003). The first condition ([7]--[15] $>$ 1.8 with
[15] $<$ 4.5, i.e. F(15$\mu$m) $\ga$ 0.3 Jy) is met by 715 7--15~$\mu$m 
sources of the ISOGAL catalogue, which are thus luminous
YSO candidates. However, they represent only 2\% of the total number of sources with good detection at both wavelengths. One should add about a hundred  objects with comparable 15~$\mu$m flux density, not
detected at 7~$\mu$m,  most of which are probably young. The majority of
these young ISOGAL stars are located in the molecular ring, or in
the inner Bulge; their projected density is considerably higher in
the latter (Felli et al. 2002). In his PhD Thesis, Schuller (2002) has made a complete analysis of the ISOGAL and MSX YSO candidates in the inner Bulge, within $\vert \ell
\vert$~$<$~1.65$^\circ$, $\vert$ {\it b} $\vert$~$<$~0.5$^\circ$. Selecting them from their red 7--15~$\mu$m colour 
and their extended 15~$\mu$m emission, he has identified several hundred of them with O star luminosities. This allows a preliminary discussion of the present global formation rate of massive stars in the inner Bulge. 

The ISOGAL catalogue includes a number of nearer, less massive young
stars, mainly in nearby spiral arms at 1 or 2 kpc. Their colour
[7]--[15] $\ge$ 1.2 is similar to the young stars in the main star
forming regions, much closer to the Sun, studied by Nordh et
al. (1996) and Bontemps et al. (2001). For many of these ISOGAL sources it is
difficult to confirm their identification. However, such
groups of sources are clearly apparent in some fields where other
evidence of star formation is known, such as M16. A complete census of
such ISOGAL low-mass young stars is still to be done.  However,
dedicated studies of particular regions are in progress.

Of course, such YSO analyses cannot be decoupled from studies of
mm/submm molecular line and dust emission, and the identification of
mid-IR dark condensations as discussed in Sect. 5.2. In some cases,
red ISOGAL point sources, possibly YSOs, seem to be associated with
such condensations (Teyssier et al. 2002).

\subsection{Foreground stars with mid--IR excess}

A small fraction (less than 1\%) of ISOGAL sources seem to display a
mid-IR excess while their visible/near-IR colours (e.g. I--J) are
relatively blue. The latter imply little or no interstellar
extinction, and thus that such sources are relatively nearby
(typically a few 10$^2$ pc), in the foreground with respect to the bulk
of the ISOGAL sources in a given field. In a number of them, the value
of I--J ($\la$ 1) points to early spectral types (B to G) which has
been confirmed by optical spectroscopy for some of them (Schultheis et
al. 2002, Schuller et al. in preparation). 

The first question to
address is the reality of the associations of the sources detected in
the different bands, especially between DENIS and ISOCAM sources. A
close inspection of the astrometric accuracy of the
cross-identification and of the near-IR
DENIS and 2MASS images shows that in many cases spurious associations
between a foreground main--sequence star and a distant AGB star may
not be excluded (Schultheis et al. 2002). However, there remains a
number of cases where there is no direct evidence to doubt the associations. Visible spectroscopy has confirmed a
few tens   of such cases of apparently good associations of 
15~$\mu$m excess objects with early spectral types. It seems unlikely that the
majority of them could be spurious associations, especially for the
large fraction ($\sim$50\%) which display emission lines. These
ISOGAL sources should thus belong in most of cases to one of the classes
of early type objects which are known to display mid-IR
excess. 

It is known that such a combination of colours can correspond to a
configuration with optically thin distant dust, such as a face--on
disk (or a detached shell, or binary objects such as dusty symbiotic stars). The most frequent cases are young stars
with such a disk, and the many cases of A or B stars with emission
lines could be mostly Herbig Ae/Be stars. It is possible that a few
other cases are older objects with debris disks such as $\beta$ Pic or
Vega. However, it will be difficult to confirm such cases without
further mid-- to far--IR observations, in particular with SIRTF or ASTRO-F,
complemented with deep near--IR imaging to confirm the reality of the
mid/near-IR associations. A last obvious possibility is post--AGB
stars or related classes of stars. However, such objects are known to
be rare so that the probability of finding nearby ones, within the
limited total area observed by ISOGAL, remains small. Finally, we
note that Schultheis et al. (2002) have found in a couple of such B
stars an intriguing bump between 5000 and 6000 $\AA$ which could be
similar to the extended red emission (ERE) known in reflection nebulae
and in the post--AGB Red Rectangle. 

\section{Summary} 
With the completion of the data reduction and processing of the
ISOGAL mid-infrared observations on about 16 sq. degrees in the Galactic
plane, and the subsequent detection of about 100000 cool stars in up to five infrared bands (ISO 7-15 $\mu$m, DENIS I, J,
K$_{\rm s}$), the ISOGAL survey provides new insights into the  stellar populations
and interstellar mid-IR diffuse emission of the Central Galaxy, and how they are different from those in the vicinity of the solar system. High
sensitivity (approaching 10-20 mJy in normal fields outside of star-forming regions) and good spatial resolution (6$\arcsec$ or better)
make the ISOGAL survey about two orders of magnitude deeper than IRAS
in the central Galactic Disk, and more sensitive than the MSX survey by one or two magnitudes at 7--8~$\mu$m, and by three or four  magnitudes at 15~$\mu$m,
(although MSX observed an area two hundred times larger, but with a pixel size three times larger). The observed fields
sample the Galactic Disk, mainly within $\vert \ell
\vert$~$<$~30$^\circ$, $\vert$ {\it b} $\vert$~$<$~1$^\circ$, mostly avoiding
regions of active star formation where bright sources could have
saturated the detectors. Special observation modes were developed
allowing the study of a few star-forming regions with data often of poorer
quality; in particular a large fraction of the area within 1$^\circ$
of the Galactic Center was  observed with 3$\arcsec$ pixels.

Version 1 of the ISOGAL Point Source Catalogue (ISOGAL-PSC-V1) is being 
released at CDS/VizieR and on the ISOGAL web site, www-isogal.iap.fr/,
simultaneously with the present paper and with the catalogue's Explanatory Supplement
(Schuller et al. 2003). The ISOGAL-PSC-V1 includes systematic
cross-identifications with DENIS data, from special DENIS observations
and data processing (Simon et al. in preparation), and detailed
quality flags. Data processing and data quality are discussed in
detail in Schuller et al. (2003). The released data have been
conservatively selected to ensure  well-defined  reliability and
photometric quality. The typical ISOGAL photometric accuracy is
generally about 0.2 mag rms. The ISOGAL astrometry is tied to DENIS
astrometry in most fields ; it is thus better than $\sim$0.5$\arcsec$ for sources
with DENIS counterparts (partly limited by the accuracy of the present
DENIS reference Catalogue, USNO\_A2).

The high spatial resolution information on diffuse mid-infrared
emission in ISOGAL images provides a wealth of information about its
carriers (PAHs and dust, Fig. 5), as well as about the interstellar radiation which ex s them. The ISOGAL survey identified a
population of narrow (down to 10$\arcsec$ in size) very dark filaments
and globules, seen in absorption in front of diffuse galactic emission
at 7 and 15 $\mu$m, which are visible in practically all ISOGAL
images. Such infrared dark clouds are similar to those detected by the
MSX survey on larger scales, as well as by IRAS with a much lower angular resolution. Their typical properties -- dimension a
fraction of a parsec, density above 10$^5$ cm$^{-3}$ and mass more
than 10$^3$ M$_\odot$ -- have been confirmed by millimetre-wave  observations. 

The very low value of the interstellar extinction at mid-IR
wavelenths means that ISOGAL data are useful complements to near-IR data in
determining extinction along lines of sight with very large
extinction, values of A$_V$ $\ga$ 20. One may use, e.g., either the
value of the  K$_{\rm s}$--[7] colour along the line of sight of DENIS or 2MASS red giants without circumstellar dust, or the dimming of
mid-IR diffuse emission by infrared dark clouds. However, such methods
rely on well--established values for the mid--IR interstellar
extinction law. Preliminary analysis of ISOGAL data tends to confirm
that this extinction law is close to classical values (see e.g. Mathis
1990) in most of the Galactic Disk. However, continuing analyses may provide evidence that it differs in certain regions such as
the vicinity of the Galactic Center.

The several colour--magnitude diagrams (CMD) and colour--colour
diagrams that one can create with the diverse combinations of the
five bands from ISOCAM and DENIS can effectively distinguish  between
intrinsic source properties and effects due to interstellar
extinction. Most ISOGAL sources are red giants with luminosities within
a couple of magnitudes of the tip of the first giant branch. Most 15
$\mu$m sources are AGB stars above the first GB tip and are 
long--period or semi--regular 
variables. ISOGAL sources are useful tracers of the stellar
populations of the central Galaxy, especially in the most obscured
regions of the inner Galactic Bulge and Disk. Multi-wavelength data
allow one to estimate the luminosity function in the inner Bulge where the
distance is known. A preliminary analysis there confirms the
presence of a relatively young component together with the Old Bulge
population.

The 15~$\mu$m excess, determined by multi-wavelength analysis and
observed in a large fraction of 15~$\mu$m ISOGAL AGB sources, allows a
systematic analysis of AGB mass-loss in the inner Galactic Disk and
Bulge from a sample of more than 10$^4$ sources. 
Mass-loss rates are determined from recent 
modeling of their circumstellar dust (Jeong et al. 2002), allowing a preliminary estimate of the total mass-loss returned to the interstellar medium. The few ISOCAM--CVF spectra
(5--16~$\mu$m) observed in the ISOGAL program will be useful for this 
purpose, together with theoretical modeling and SIRTF or ASTRO-F
spectra.

ISOGAL has also detected a number of young stars of diverse types with
circumstellar dust. They range from massive YSOs at large distances,  up
to that of the Galactic Center, to less massive stars mainly in nearby spiral
arms at 1 or 2 kpc. Analysis is in progress from ISOGAL colours
and additional information.

In addition to ``normal'' evolved and young stars, the very large ISOGAL
sample also contains a variety of peculiar stars with circumstellar
dust : supergiants, post-AGB stars, planetary nebulae, symbiotics,
foreground early type stars, etc., with in particular diverse types of
circumstellar disks. The ISOGAL data are thus a treasury, still little
exploited, for follow-up observations especially by future mid-IR
space missions such as SIRTF, ASTRO-F and NGST, and high--resolution
ground--based studies, aimed at the circumstellar physics and evolution of
such objects.

The present results derived from the ISOGAL survey are published or in
advanced preparation in about 25 refereed papers. We are already
considering a second release of ISOGAL products with improved quality and
with extended associations : better-corrected images
(Miville--Desch\^enes et al. 2000 and in preparation); a V2 point source
catalogue with better photometry and sensitivity;
associations with 2MASS, MSX and IRAS sources ; better astrometry with
new DENIS astrometry and 2MASS, etc.

\medskip

{\bf Acknowledgements}\\ 
The ISOCAM data presented in this paper were analysed using `CIA', a joint development by the ESA Astrophysics Division and the ISOCAM Consortium. The ISOCAM Consortium is led by the ISOCAM PI, C. Cesarsky. We thank T. Prusti, R. Siebenmorgen, H. Aussel, R. Gastaud, J.L. Starck and many other members of the ISOCAM team, of the ISO/ESA team at Villafranca and especially of the CIA team for their constant help in the ISOGAL observations and data reduction. We thank the ISO/SWS team and its PI, T. de Graauw, for providing the ISOCAM data of the GPSURVEY project for a joint analysis with ISOGAL data.

This work was carried out in the context of EARA, the European Association  for Research in Astronomy. 

S. Ganesh and D. Ojha were supported by a fellowship from the
Minist\`ere des Affaires Etrang\`eres, France, and this project was
supported by the Project 1910-1 of the Indo-French Center for the
Promotion of Advanced Research (CEFIPRA), as well as by CNES. M. Schultheis acknowledges
the receipt of an ESA fellowship. B. Aracil, T. August, X. Bertou,
P. Hennebelle and A. Soive were posted to the ISOGAL Project by the
D\'el\'egation G\'en\'erale de l'Armement, France. The participation of I. Glass in the ISOGAL project was
supported by a CNRS-NRF grant (NRF= National Research Foundation, South Africa).
 
We thank all the members of the DENIS team who allowed obtaining the DENIS data. The DENIS project is partially funded by European Commission through
SCIENCE and Human Capital and Mobility plan grants. It is also
supported, in France by the Institut National des Sciences de
l'Univers, the Education Ministry and the Centre National de la
Recherche Scientifique, in Germany by the State of Baden-W\"urtemberg,
in Spain by the DG1CYT, in Italy by the Consiglio Nazionale delle
Ricerche, in Austria by the Fonds zur F\"orderung der
wissenschaftlichen Forschung und Bundesministerium f\"ur Wissenshaft
und Forschung, in Brazil by the Foundation for the development of
Scientific Research of the State of Sao Paulo (FAPESP), and in Hungary
by an OTKA grant and an ESOC\&EE grant.

\bibliographystyle{aastex}

\end{document}